\documentclass[twocolumn]{IEEEtran}

\usepackage{cite}
\usepackage[Gray,amssymb]{SIunits}
\usepackage{color}
\usepackage{acronym}
\usepackage{esvect}
\usepackage[hidelinks]{hyperref}
\usepackage{amsfonts,amsmath,amssymb}
\usepackage{dsfont}
\usepackage{nicefrac}
\usepackage{multirow}
\usepackage{booktabs}
\usepackage{bbm}
 \usepackage[table]{xcolor}
\ifCLASSINFOpdf
  \usepackage[pdftex]{graphicx}
  \DeclareGraphicsExtensions{.pdf,.jpeg,.png}
\else
  \usepackage[dvips]{graphicx}
  \DeclareGraphicsExtensions{.eps}
\fi
\usepackage{datetime}
\usepackage[makeroom]{cancel}

\allowdisplaybreaks

\definecolor{orange}{rgb}{1,0.5,0}

\newcommand{\M}{{\mathcal{M}}}
\newcommand{\D}{{\mathcal{D}}}

\newcommand{\R}{\mathbb{R}}
\newcommand{\T}{\mathrm{T}}
\renewcommand\vec[1]{\boldsymbol{#1}}%

\everymath{\displaystyle}

\newcommand{\eye}{\ensuremath{\mathbbm{1}}}
\newcommand{\oo}{\ensuremath{\mathbbm{O}}}

\acrodef{hvdc}[\textsc{hvdc}]{high voltage direct current}
\acrodef{mtdc}[\textsc{mtdc}]{multi-terminal direct current}
\acrodef{pfs}[\textsc{pfs}]{power flow solution}
\acrodef{pcc}[\textsc{pcc}]{point of common coupling}
\acrodef{poc}[\textsc{poc}]{point of connection}
\acrodef{opf}[\textsc{opf}]{optimal power flow}
\acrodef{hb}[\textsc{hb}]{harmonic balance}
\acrodef{smps}[\textsc{smps}]{switched mode power supply}
\acrodef{mvf}[\textsc{mvf}]{multi-variate formulation}
\acrodefplural{smps}[\textsc{smps}]{switched mode power supplies}
\acrodef{pll}[\textsc{pll}]{phase locked loop}
\acrodef{sh}[\textsc{sh}]{shooting}
\acrodef{dae}[\textsc{dae}]{differential algebraic equation}
\acrodef{ode}[\textsc{ode}]{ordinary differential equation}
\acrodef{sde}[\textsc{sde}]{stochastic differential equation}
\acrodef{efm}[\textsc{efm}]{envelope-following method}
\acrodef{bvp}[\textsc{bvp}]{boundary value problem}
\acrodef{lte}[\textsc{lte}]{local truncation error}
\acrodef{lcm}[\textsc{lcm}]{least common multiple}
\acrodef{rmse}[\textsc{rmse}]{root mean squared error}
\acrodef{vc}[\textsc{vc}]{virtual connector}
\acrodef{ps}[\textsc{pss}]{power electromechanical sub-system}
\acrodef{ess}[\textsc{ess}]{electromagnetic sub-system}
\acrodef{mna}[\textsc{mna}]{modified nodal analysis}
\acrodef{psm}[\textsc{psm}]{power system model}
\acrodef{emt}[\textsc{emt}]{electromagnetic transient}
\acrodef{ts}[\textsc{ts}]{transient stability}
\acrodef{shpf}[\textsc{shpf}]{shooting power flow}
\acrodef{pf}[\textsc{pf}]{power flow}
\acrodef{vsc}[\textsc{vsc}]{voltage source converter}
\acrodef{mppt}[\textsc{mppt}]{maximum power point tracker}
\acrodef{pic}[\textsc{pi}]{proportional-integral}
\acrodef{ibr}[\textsc{ibr}]{inverter-based resource}
\acrodef{hvdc}[\textsc{hvdc}]{high-voltage direct current}
\acrodef{mtdc}[\textsc{mtdc}]{multi-terminal direct current}
\acrodef{iqr}[\textsc{iqr}]{interquartile range}
\acrodef{ou}[\textsc{ou}]{Ornstein-Uhlenbeck}
\acrodef{psd}[\textsc{psd}]{power spectral density}
\acrodef{coi}[\textsc{coi}]{center of inertia}
\acrodef{cig}[\textsc{cig}]{converter-interfaced generator}
\acrodef{gf}[\textsc{gf}]{grid-forming}
\acrodef{vf}[\textsc{vf}]{vector fitting}
\acrodef{snr}[\textsc{snr}]{signal-to-noise-ratio}

\newcommand{\InRefFig}[1]{Figure~\ref{#1}}

\newcommand{\RefFig}[1]{Fig.~\ref{#1}}

\newcommand{\InRefEq}[1]{Equation~(\ref{#1})}

\newcommand{\RefEq}[1]{equation~(\ref{#1})}

\newcommand{\RefE}[1]{(\ref{#1})}

\newcommand{\RefSec}[1]{Sec.~\ref{#1}}

\begin{document}

\title{Global Momentum Estimation of\\ an Electric Power System}

\author{Angelo Maurizio Brambilla, \IEEEmembership{Senior Member,~IEEE},
  Davide del Giudice, \IEEEmembership{Student Member,~IEEE}, Daniele Linaro,
  \IEEEmembership{Member,~IEEE}, and Federico Bizzarri,
  \IEEEmembership{Senior Member,~IEEE}\thanks{F.~Bizzarri is with Politecnico
    di Milano, DEIB, p.zza Leonardo da Vinci, no. 32, 20133 Milano,
    Italy and also with the Advanced Research Center on Electronic
    Systems for Information and Communication Technologies E.~De
    Castro (ARCES), University of Bologna, 41026 Bologna, Italy.
    (e-mail: federico.bizzarri@polimi.it).} \thanks{D.~del Giudice,
    D.~Linaro, and A.~Brambilla are with Politecnico di
    Milano, DEIB, p.zza Leonardo da Vinci, no.~32, Milano, 20133,
    Italy.  (e-mails: \{davide.delgiudice, samuele.grillo,
    daniele.linaro, angelo.brambilla\}@polimi.it).}}

\maketitle

\begin{abstract}
This work presents a technique to estimate on-line the global momentum 
of an electric power system.
It exploits the footprint of the principal frequency system dynamics.
Probing tones are injected by a grid-forming converter-interfaced generator
and the speed of its virtual rotor is exploited to extrapolate the footprint.
The proposed technique is entirely data driven and thus it does not need 
any model of the power system.
We show that the proposed technique is adequate to accurately estimate 
the actual global momentum contributed by synchronous machines and 
provided by the controllers of converter-interfaced generators 
that emulate the behavior of synchronous machines.
The technique is comprehensively tested on a modified version of the 
\textsc{ieee 39-bus} system and a dynamic version of the \textsc{ieee 118-bus} system,
containing grid forming converter interfaced generators.
\end{abstract}


\vspace{-2mm}
\section{Introduction}

\subsection{Motivation}
From the beginning of this century, the importance of estimating an
electric power grid's available inertia---or better yet, its
momentum---has been steadily increasing. To a large extent, this
explains the large amount of papers on this topic that one can find in
the scientific literature.
The interested reader can find the motivations supporting this trend 
in the introductions of those contributions and, in particular, in extensive 
review papers such as \cite{KONTIS:2021,PRABHAKAR:2022}.

\subsection{Literature Review}
The literature concerning inertia estimation in power systems 
is vast and has been growing at a quickening pace. 
Limited to the ``Scopus'' abstract and citation database of peer-reviewed 
papers, papers including the keywords ``inertia'' and ``power system''
in their abstract were only $30$ in $2000$, while this figure spiked to $1200$ in $2022$:
nevertheless, the problem is still open, despite the untiring work of numerous
research groups.

In this context, we present a reliable and efficient method for 
estimating online the {\em global momentum} available in a power system.
The method belongs to the class of algorithms based on {\em active
  perturbations}, where proper probing signals\cite{5338001} are used
for system identification purposes\cite{Zhou:2006,Chakraborty:2022}.

Focusing on this class of approaches, the method proposed in 
\cite{Hosaka:2019} was verified in an actual power system and resulted 
effective for small-scale islanded systems. 
In \cite{Tamrakar:2020}, a power system was probed by small active 
power changes through an energy storage system that did not impact the 
operational stability of the system. 
The goal was to provide the energy storage system operators with a tool 
to estimate the unknown time-varying inertia of any generic power system 
and properly tune/control their devices for fast-frequency strategies. 
Low-level probing signals were used also in \cite{Rauniyar:2021} for 
accurate estimation of inertia and damping constants in microgrids by 
resorting to a moving horizon estimation approach.
The method proposed in \cite{7913675} is possibly the most well-known in 
the considered class. 
The authors proposed a closed-loop identification technique to estimate 
the equivalent inertia constant of a power system at the connection bus.
The method aims to estimate the inertia constant of a single device 
connected to the grid.

The method we propose has a wider scope since it allows to estimate online
the global momentum of large-size realistic power system. 
The main source of inspiration for our work was the activity conducted by other researchers in the realm of
frequency synchronization of power generators. The latter  is a necessary condition for the
operation of power-grid networks. 
During steady operation, the frequency is the same throughout the entire
power grid and any pair of generators has a fixed phase difference 
that determines the power flows.
Even if the focus of our work is not synchronization but global momentum 
estimation, the main results we propose are grounded on this concept.
It is thus worth mentioning that one can find in the literature many 
contributions on synchronization (see for instance 
\cite{Sajadi2022, Gorjao2022, Guo2021, Zhu2018, Nishikawa2015, 
Motter2013,Ramaswamy1995}).

\subsection{Contribution}
Here, we provide a theoretical framework based on the
dynamic response of a generic power system model containing synchronous 
generators equipped with primary frequency control and \ac{gf} \acp{cig} 
whose swing equations are described by the classical model  \cite{Kundur1994,9625933}.
By interpreting the power system as a power-controlled oscillator whose instantaneous
oscillation frequency is controlled by the power flows, we derive the
analytical expression in the frequency domain of the {\em principal 
frequency system dynamics}. We show that this expression represents the behavior shared by the rotor speed of all the synchronous generators and thus also by \ac{gf} \ac{cig} elements
present in the grid, provided that a proper amount of 
{\em synchronization} (at low frequencies) is guaranteed between these 
devices. One of the parameters of this expression
is the global momentum of the grid, which can thus be estimated with
appropriate identification techniques.
We show that the estimation depends neither on the electrical characteristics
of the interconnecting lines, nor on the number of buses, controllers, governors, 
and synchronous machine models but only on the collected data, i.e., 
the proposed estimation method is agnostic with respect to the system.

Based on these considerations, we exploit a \ac{gf} \ac{cig} that
is either already present in the power system, or can be inserted with
the specific purpose of allowing the estimation of the power system
global momentum. The probing signals used to stimulate the power
system in a suitable frequency band are injected by the \ac{gf}
\ac{cig} and its virtual rotor speed is sampled at an appropriate rate.
By resorting to the \ac{vf} technique \cite{772353,1645204,Triverio},
we obtain the parameters that provide the best fit of the analytical 
expression of the frequency spectrum of the collected samples. The global momentum of the power system is among these parameters.

\subsection{Organization}
In \RefSec{S:standard} we present the theoretical foundation of the 
proposed method. 
In \RefSec{S:insight}, through a basic case-study, we highlight how
the global momentum of the power system can be identified in the 
low-frequency spectrum expression of the rotor speed of all the synchronous 
generators and \ac{gf} \ac{cig} elements that are present in the grid. 
The general formulation of the method is given in \RefSec{S:method}.
Finally, in \RefSec{S:NumExp} we verify the validity of the 
proposed approach in the \textsc{ieee 39-bus} and \textsc{ieee 118-bus} 
power system models.

\section{The Power System as a Power-Controlled Oscillator}
\label{S:standard}
In electronics, a voltage-controlled oscillator is an oscillator whose
oscillation frequency is controlled by a voltage input: the applied
voltage determines the instantaneous oscillation
frequency. Analogously, a generic power system can be viewed as a
power-controlled oscillator whose instantaneous oscillation frequency,
i.e., rotor speed of synchronous generators, is controlled by the power flows.

In an autonomous oscillator, steady-state periodic solutions lack a 
phase reference \cite{Farkas1994,Kuznetsov2004}. Even if a limit cycle is 
unique and isolated in the phase space, two orbits originating from two 
different initial conditions belonging to this limit cycle will remain 
shifted in the time domain. This property is at the origin of phase noise in 
electronic oscillators \cite{Demir2000}. From the viewpoint  of  dynamical 
systems theory, according to Floquet's theory, this is justified by the 
presence of a characteristic multiplier equal to $1$. 

The same holds in power systems where a stationary solution is not an {\em
isolated  equilibrium} but is embedded in a {\em continuum of
equilibria} \cite{Aulbach1984}. When the power system is modeled in
the dq-frame, the presence of a characteristic multiplier equal to $1$
corresponds to a eigenvalue equal to $0$ in the Jacobian matrix of the power
system model linearized around an equilibrium point \cite{Sauer1990}.
Indeed, the classical power system model aims at representing
the  envelope of the {\em actual} system dynamics through a
steady-state solution. In other words, the {\em periodic steady-state
solution} at the fundamental frequency with a {\em constant}
envelope is actually represented as a {\em constant} steady-state
solution playing the role of a {\em stationary solution} in the dq-frame.

In reality, the power system model is more complex than a voltage-controlled 
oscillator since it is made up of several power-controlled oscillators (viz., 
synchronous generators and \ac{gf} \acp{cig}) \cite{Sajadi2022,Motter2013}. 
An overall instantaneous oscillation frequency, shared by all of these 
components, is observed only when these local and interconnected oscillators 
are synchronized. In practice, even in normal operating conditions, it never 
occurs but {\em on average}. As a matter of fact, the power system
model never works at steady state because of random fluctuations in 
the power consumption of the loads or in the energy production of renewable 
energy sources.

To clarify these concepts, it is useful to discuss what happens in a 
power system model including $N$ synchronous generators
whose dynamic evolution is modeled by the simple swing equation\footnote{
The extension to the case in which \acp{cig} are present and
modeled by the simple swing equation is straightforward.} and equipped with
primary frequency control.
Generators are interconnected by lines and transformers, and constant-power 
and/or constant-impedance loads are connected to the grid buses. Under these 
assumptions, the overall power system model is described by the 
following set of $3N$ \acp{ode},
\begin{equation}
  \label{E:FullSys}
  \begin{aligned}
    {\vec{\dot \delta}} &= \Omega \left(\vec \omega - \omega_0 \right)\\
    \vec \M \vec{\dot \omega} &=
    \vec P_{m} -
    \vec P_e(\vec \delta)
    - \vec \D (\vec \omega-\omega_0)\\
		\vec T_{g} \vec{\dot{P}}_m &=
    \vec P_{{m_{\mathrm{eq}}}}-\vec P_{m} - {\vec k_{g}}{\vec R^{-1}}(\vec \omega-\omega_0)~,
  \end{aligned}
\end{equation}
where the meaning of symbols in \RefE{E:FullSys} is as follows:
\begin{itemize}
\item[--] $\Omega$: the base synchronous frequency in
  $\rad/\second$;
\item[--] $\vec \omega(t) \in \mathds{R}^{N}$: the per-unit rotor speeds of
  the machines;
\item[--] $\omega_0 \in \mathds{R}$: the per-unit reference synchronous
  frequency;
\item[--] $\vec \delta(t) \in \mathds{R}^{N}$: the rotor angles of
  the machines;
\item[--] ${\vec \M \in \mathds{R}^{N \times N}}$: a diagonal matrix
  whose $j$-th element is the product of the inertia constant $H_{j}$
  and the rated power $S_{B_{j}}$ of the $j$-th machine:
  $\mathcal{M}_{jj}=2H_{j}S_{B_{j}}$ for ${j=1,\dots,N}$\footnote{$\mathcal{M}_{jj}$ is a scaled version
    of the generator momentum given by ${4\pi\,\Omega^{-1}H_{j}S_{B_{j}}}$.
    The coefficient ${2\pi\,\Omega^{-1}}$ is not present since $\vec\omega$
    is expressed in per-unit. Nevertheless, in the following, we will
    refer to $\mathcal{M}_{jj}$ as the generator momentum.};
\item[--] ${\vec \D \in \mathds{R}^{N \times N}}$: a diagonal matrix
  whose $j$-th element is the product of the load damping factor $D_j$
  and the rated power $S_{B_{j}}$ of the $j$-th machine: $\D_{jj}=D_jS_{B_j}$
  for ${j=1,\dots,N}$;
\item[--]
  $\vec P_e(\vec \delta) \in
  \mathds{R}^{N}$: the electrical active power exchanged by the machines;
\item[--] ${\vec P_m} \in
  \mathds{R}^{N}$: the mechanical power of the machines, governed by their prime mover;
\item[--] ${\vec P_{m_{\mathrm{eq}}}} \in
  \mathds{R}^{N}$: the mechanical power setpoint of the machines corresponding to the 
  power flow solution;
\item[--] ${\vec T_{g}}$, ${\vec k_{g}}$,  ${\vec R} \in \mathds{R}^{N \times N}$:
the diagonal matrices whose elements model the time constant, gain constant, and droop gain of
the primary frequency controls, respectively\footnote{
For the sake of simplicity, we assume that all the $N$ synchronous generators
are equipped with a primary frequency control. The less generic case in which
only a subset of them shares this property can be straightforwardly formalized.}.
\end{itemize}
It is worth noticing that the first set of $N$ \acp{ode} in \RefE{E:FullSys}
should be written as ${\vec{\dot \delta}} = \Omega \vec \omega$ if one wanted to completely retain
the formulation in polar coordinates of the synchronous-generator dynamical equations, in which the
rotor angle grows unbounded since the rotor keeps rotating.
Nevertheless, \RefE{E:FullSys} is usually adopted to identify an equilibrium
point of the system, say $[\vec \delta_{\mathrm{eq}}^\T, \vec \omega_{\mathrm{eq}}^\T,\vec P_{m_{\mathrm{eq}}}^\T]^\T$, neglecting the common shifting evolution of the ${\vec \delta}$
rotor angles.

Given a common shift of the ${\vec \delta}$ rotor angles written as
${\vec\delta(t)+\alpha(t)}$, where $\alpha(t)$ is a scalar function, 
we have that ${\vec P_e(\vec \delta(t)+\alpha(t)) = \vec P_e(\vec \delta(t))}$.
As a consequence\footnote{In the following $\vec \eye_k$ is the 
${k \times k}$ identity matrix, $\vec \oo_k$ is a ${k \times k}$ matrix 
of zero elements, and $\vec \eye_{k,h}$ and $\vec \oo_{k,h}$ are 
${k \times h}$ matrices of $1$ or $0$, respectively.}, by considering 
the linearisation of \RefE{E:FullSys} we have
\begin{equation}
  \label{E:FullSysLin}
	\left[\!\!
  \begin{array}{c}
	\Delta \vec{\dot \delta}\\
	\Delta \vec{\dot \omega}\\
	\Delta \vec{\dot{P}}_m\\
  \end{array}
	\!\!\right]\!\!
	= \!\!
	\underbrace{
	\left[\!\!\!\!
  \begin{array}{c c c}
	\vec \oo_N & \Omega \vec \eye_N & \vec \oo_N\\
	-\vec \M^{-1} \vec \Pi & -\vec \M^{-1} \vec \D & \vec \M^{-1}\\
	\vec \oo_N & -\vec T^{-1}_{g} \vec k_{g} \vec R^{-1} & -\vec T^{-1}_{g}\\
	\end{array}
	\!\!\!\!\right]}_{\vec A}
	\!\!
	\left[\!\!
  \begin{array}{c}
	\Delta \vec{\delta}\\
	\Delta \vec{\omega}\\
	\Delta \vec{P}_m\\
	\end{array}
	\!\!\right]~,
\end{equation}
where the $\vec \Pi =\frac{\partial{\vec P_e}}{\partial \vec{\delta}}$ 
matrix is the network interconnection Laplacian {\em singular}
matrix \cite{Machowski2020} and $\vec \Pi= \vec \Pi^\T$.
The null space of $\vec \Pi$ is spanned by $\ker(\vec \Pi)=\vec \eye_{N,1}$. 
The $\vec A$ matrix is thus singular too and 
$\ker(\vec A)\equiv \vec u_1=[\vec \eye_{1,N}, \vec \oo_{1,2N}]^{\T}$.

When a small-signal ${\vec b(t) \in \R^N}$ is added to 
$\vec P_e(\vec \delta(t))$, thus emulating the injection of (stochastic) disturbances
or probing signals, the $N$ \acp{ode} in \RefE{E:FullSys} governing 
${\vec{\dot \omega}}$ become
\begin{equation}
  \label{E:HalfSysNoise}
    \vec \M \vec{\dot \omega} =
    \vec P_m -
    \vec P_e(\vec \delta)
    - \vec \D (\vec \omega-\omega_0) + \vec b(t) ~.
\end{equation}
Since $\vec A$ is singular, it is not possible to exploit \RefE{E:FullSysLin} 
to obtain an approximate solution of \RefE{E:HalfSysNoise} in the neighborhood
of $[\vec \delta_{\mathrm{eq}}^\T, \vec \omega_{\mathrm{eq}}^\T, 
\vec P_{m_{\mathrm{eq}}}^\T]^\T$. 
Nevertheless, in this case, the singularity of $\vec \Pi$ is a key aspect to allow resorting to the small-signal formulation to study the perturbed evolution of $\vec \omega(t) = \vec \omega_{\mathrm{eq}} + \Delta \vec \omega(t)$. 
To explore this aspect, let us define $\vec u_k$ and $\vec v_k$ as the 
$3N$ right and left eigenvectors of $\vec A$, respectively
($\vec u_1$ and $\vec v_1$ are associated to the $\lambda_1=0$ eigenvalue).
So doing we are assuming that $\vec A$ has distinct eigenvalues (their value
is related to the eigenvalues of $\vec \Pi$ \cite{Machowski2020} and to the
primary frequency control time-constants) and it is diagonalizable.

Having in mind that, because of the bi-orthogonality property of eigenvectors (i.e., ${\vec v_k^\T \vec u_j \neq 0}$ only if $k \neq j$), it is always possible 
to write\footnote{In general, for $k=1 \dots 3N$, 
${\vec v_k^\T \vec u_k \neq 1}$ but it is always possible to scale each 
${\vec v_k}$ w.r.t. ${\vec v_k^\T \vec u_k}$, thus obtaining a new vector 
${\hat{\vec v}_k}$ such that ${{\hat{\vec v}_k}^\T \vec u_k = 1}$. 
This is done in the following but omitting the $\hat{}$ symbol
to keep the notation terse.}
\begin{equation}
  \label{E:DecompNoise}
	\left[\!\!\!\!
	\begin{array}{c}
	  \vec \oo_{N,1}\\
    \vec \M^{-1} \vec b(t)\\
		\vec \oo_{N,1}
	\end{array}	
	\!\!\!\!\right]	
		= \underbrace{\vec v_1^\T
	\left[\!\!\!\!
	\begin{array}{c}
	  \vec \oo_{N,1}\\
    \vec \M^{-1} \vec b(t)\\
		\vec \oo_{N,1}
	\end{array}	
	\!\!\!\!\right]	
	\vec u_1}_{\vec b_\delta(t)} + \underbrace{\sum_{k=2}^{3N}{\vec v_k^\T
	\left[\!\!\!\!
	\begin{array}{c}
	  \vec \oo_{N,1}\\
    \vec \M^{-1} \vec b(t)\\
		\vec \oo_{N,1}
	\end{array}	
	\!\!\!\!\right]	
	\vec u_k}}_{\vec b_{\delta \omega}(t)}~.
\end{equation}
Since
\begin{equation}
  \label{E:v1}
	\vec v_1=\ker(\vec A^\T)=\frac{\Omega}{\vec \eye_{1,N} \vec \Theta \vec \eye_{N,1}}
	\left[
	\begin{array}{c}
	{\vec \eye_{N,1} \vec \Theta}{\Omega^{-1}}\\
	  \vec \M \vec \eye_{N,1}\\
		\vec T_{tg} \vec \eye_{N,1}
	\end{array}
	\right]~,
	\end{equation}
one obtains
\begin{equation}
  \label{E:bdelta}
	\vec b_\delta(t) = \frac{\Omega}{\vec \eye_{1,N} \vec \Theta \vec \eye_{N,1}}
	\left[
	\begin{array}{c}
	\vec \eye_{1,N} \vec b(t) \vec \eye_{N,1}\\
	  \vec \oo_{2N,1}
	\end{array}	
	\right]~,
\end{equation}
where $\vec \Theta = \vec \D+\vec k_{g} \vec R_{g}^{-1}$ (see Appendix A for more
details on the derivation of $\vec v_1$).

The effect of ${\vec b_\delta(t)}$ is thus to {\em synchronously} shift all the components of
${\vec \delta}$ of the same time-varying quantity
\begin{equation}
\label{E:alpha}
	\alpha(t) = \Omega \int_0^t{\frac{\vec \eye_{1,N} \vec b(\tau)}{\vec \eye_{1,N} \vec \Theta \vec \eye_{N,1}}d\tau}~, 
\end{equation}
without altering the ${\vec \omega}(t)$ vector since
${\vec P_e(\vec \delta(t)+\alpha(t)) = \vec P_e(\vec \delta(t))}$. 
If $\eye_{1,N} \vec b(\tau)$ has a nonzero mean value $\alpha(t)$ would increase unbounded. 
This would be true even if $\vec b(\tau)$ were a 
vector of random variables with zero mean and finite variance as the \ac{ou} 
processes typically used to model stochastic variability of power 
loads \cite{Nwankpa1990, Milano2013, Hirpara2015, 7540898}.
As a matter of fact, $\alpha(t)$
would exhibit unbounded variance \cite{Abundo2013}.
In both cases, the effect of ${\vec b_\delta(t)}$
can not be treated as that of a small signal, even being ${\vec b_\delta(t)}$ actually small.
In the literature this is known as phase noise \cite{Demir2000}.

In turn, it is possible to compute the effect of
${\vec b_{\delta\omega}(t)}$ by resorting to the small-signal approach since
it is easy to verify that $\vec v_1^\T \vec b_{\delta\omega}(t) \vec u_1 = 0$. 
This implies that ${\vec b_{\delta\omega}(t)}$ does not produce any effect 
similar to $\alpha(t)$. 
In other words, it is not responsible for coherent phase shifting in the
entire power system but produces 
small fluctuations that are different for all the components 
of $\vec{\delta}(t)$. 
The main implication of this result is that, in the presence of a small signal 
$\vec b_{\delta\omega}(t)$, {\em the power system does not remain synchronized}.
In other words, it is possible to prove that, in the presence of the small signal
$\vec b_{\delta\omega}(t)$, the hypothesis that the power 
system remain synchronized, viz. $\Delta \omega_k(t) \equiv \Delta\omega(t)$
for ${k=1,\dots,N}$ (or, equivalently, ${\vec \Delta \omega(t) = \Delta \omega(t) \vec \eye_{N,1}}$)
is inconsistent with the equations governing the power system
itself. This can be derived by focusing on the \acp{ode} governing the
dynamics of $\Delta\vec{\omega}$. We can write
\begin{equation}
  \label{E:HalfSysNoise2}
	\begin{array}{l c l}
    \vec \M \Delta \vec{\dot \omega} &\!\!\! = \!\!\!&
    \Delta \vec P_m  -\vec \Pi (\alpha(t)\vec \eye_{N,1}+\Delta \vec \delta(t))
    - \vec \D \Delta \vec \omega  + \vec b(t)\\[2mm]
		  &\!\!\! = \!\!\!&
    \Delta \vec P_m  -\vec \Pi\Delta \vec \delta(t)
    - \vec \D \Delta \vec \omega + \vec \M \vec b_{\delta\omega}(t)\\[2mm]
		&\!\!\! = \!\!\!&
    \Delta \vec P_m  -\vec \Pi \int{\Omega \Delta \vec \omega(t) dt}
    - \vec \D \Delta \vec \omega + \vec \M \vec b_{\delta\omega}(t)~,
		\end{array} 
\end{equation}
with $\Delta \vec{\omega}$ evolving in the neighborhood
of $\vec \omega_{\mathrm{eq}}$, i.e., of the frequency at the equilibrium
point. 
It is worth noticing that, thanks to the singularity of $\vec \Pi$, the
possibly unbounded contribution of $\alpha(t)\vec\eye_{N,1}$ does not 
affect $\Delta \vec {\dot\omega}$.

Assuming ${\Delta \vec \omega(t) = \Delta \omega(t) \vec \eye_{N,1}}$,
${\Delta \omega(0) = 0}$, and ${\Delta \vec P_m(0) = \vec \oo_{N,1}}$, 
\RefEq{E:HalfSysNoise2} can be transformed in the $s$-domain as
\begin{equation}
  \label{E:HalfSysNoise1S}
	\begin{array}{l c l}
    s {\Delta {\omega}(s)} \vec \M \vec \eye_{N,1} \!\!\!&\!\!= \!\!&\!\!\!
        \Delta \vec P_m(s) - \Omega \frac{\Delta \omega(s)}{s} \underbrace{\vec \Pi  \vec \eye_{N,1}}_{\vec \oo_{N,1}} +
				\\[3mm]
				\!\!\!&\!\!  \!\!&\!\!\! 
    - {\Delta {\omega}(s)} \vec \D \vec \eye_{N,1} + \vec \M \vec b_{\delta\omega}(s)\\[3mm]
		  \!\!\!&\!\!= \!\!&\!\!\!
        \Delta \vec P_m(s)- {\Delta {\omega}(s)} \vec \D \vec \eye_{N,1} + \vec \M \vec b_{\delta\omega}(s) ~.
		\end{array}
\end{equation}
Since
\begin{equation}
  \label{E:Pms}
	\Delta \vec P_m(s) = -\left(s\vec T_{g}+\vec \eye_N\right)^{-1}\vec k_{g} \vec R^{-1} \vec \eye_{N,1}\Delta \omega(s) ~,
\end{equation}
we have
\begin{equation}
  \label{E:HalfSysNoise2S}
		\vec \Xi \vec \eye_{N,1} \Delta {\omega}(s)  =\vec \M\vec b_{\delta\omega}(s)~,
\end{equation}
where ${\vec\Xi=s \vec \M + \vec \D +\left(s\vec T_{g} + 
\vec \eye_{N}\right)^{-1}\vec k_{g} \vec R^{-1}}$.
Being $\vec\Xi$ a diagonal matrix that, generically, does not contain 
identical elements in its diagonal, it is not possible to have
${\Delta \vec  \omega(t) = \Delta \omega(t) \vec \eye_{N,1}}$, thus 
{\em violating the initial hypothesis}.

How is it possible to reconcile this result with the well-known 
experimental evidence that, at low-frequency, in properly connected 
power systems, the components of ${\Delta \vec \omega(t)}$ share 
{\em almost the same} frequency spectrum? 
Let us consider the small perturbation of the so-called 
{\em principal frequency system dynamics}, viz. the frequency that can be defined 
for the \ac{coi} of the system, i.e.,
\begin{equation}
	\label{E:omega_COI}
	\Delta \omega_{\textsc{coi}}(t) = \frac{\sum_{k=1}^{N}{S_{B_k}H_{k} \Delta \omega_k(t)}}{\sum_{k=1}^{N}{S_{B_k}H_{k}}}~.
\end{equation}
The \ac{ode} governing the evolution of ${\Delta \omega_{\textsc{coi}}(t)}$ is 
derived by combining the \acp{ode} of the 
$\Delta \omega_k(t)$ as \cite{Kundur1994}
\begin{equation}
\label{E:ODECOI}
\mathcal G_\M \Delta {\dot \omega}_{\textsc{coi}}=\\
\vec \eye_{1,N} \left( \Delta \vec P_{m} -
    \Delta \vec P_e
    - \vec \D \Delta \vec  {\omega} + \vec \M \vec b_{\delta\omega}(t)\right)
\end{equation}
where
$\mathcal G_\M = \vec \eye_{1,N} \vec \M \vec \eye_{N,1}$.

If we write $\vec b_{\delta\omega} = \vec b_{\delta\omega}^{\mathrm{lf}} + 
\vec b_{\delta\omega}^{\mathrm{hf}}$, where $\mbox{}^{\mathrm{lf}}$ and 
$\mbox{}^{\mathrm{hf}}$ stand for low- and high-frequency, respectively, 
and $\vec b_{\delta\omega}^{\mathrm{lf}}$ is assumed to generate 
${\Delta \vec \omega(t) \approx \Delta \omega(t) \vec \eye_{N,1}}$, we obtain
$\Delta \omega_{\textsc{coi}}(t) \approx \Delta \omega(t)$.
In the Laplace domain its dynamics is ruled by
\begin{equation}
  \label{E:HalfSysNoise2SF}
		    {\Delta {\omega}(s)} = \frac{\vec \eye_{1,N}\vec \M\vec b_{\delta\omega}^{\mathrm{lf}}(s) }{
    s \mathcal G_\M + \vec \eye_{1,N}  \left(\vec \D \vec + \left(
    s\vec T_{g} + \vec \eye_{N}\right)^{-1}\vec k_{g} \vec R^{-1} \right)\vec \eye_{N,1}}~,
\end{equation}
that is obtained, according to \RefEq{E:ODECOI}, by summing up the $N$ equations in 
\RefEq{E:HalfSysNoise2S}.
Equation \RefE{E:HalfSysNoise2SF} provides the expression of the frequency 
spectrum almost shared by all the $\Delta \omega_k(t)$ variables
at low frequency.

\section{Global momentum estimation: an insight}
\label{S:insight}
%%%%%%%%%%% BEGIN FIGURE %%%%%%%%%%%%
\begin{figure}[t!!!]
\begin{center}
\includegraphics[scale=1.0]{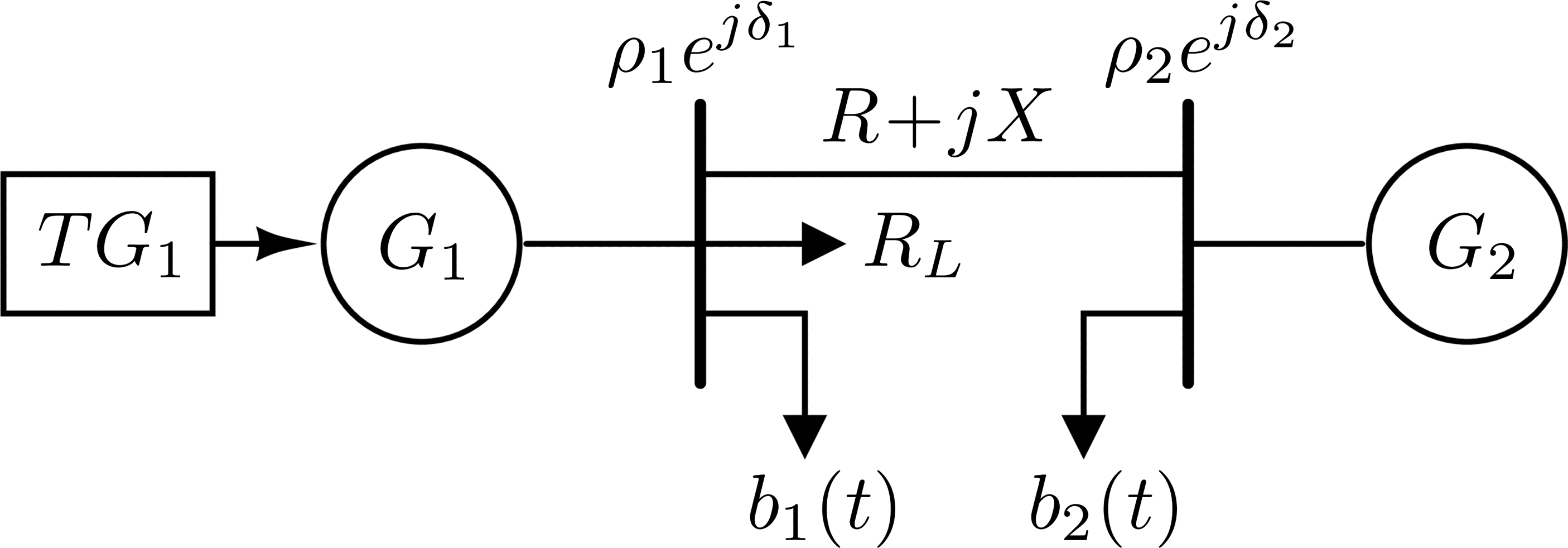}
\end{center}
\caption{The schematic of the power test system. 
$R_L$ is a resistive load, $b_1(t)$ and $b_2(t)$ are small-signal (stochastic)
perturbations that vary the active power absorbed by the grid at the power flow solution and
which are neglected in the large-signal model of the power system. For the sake of simplicity,
the internal impedance of both $G_1$ and $G_2$ is assumed to be zero, and only
the $G_1$ generator is equipped with a turbine governor.
\label{F:twomac}}
\end{figure}
%%%%%%%%%%% END  FIGURE %%%%%%%%%%%%
Our aim is exploiting \RefE{E:HalfSysNoise2SF} to systematically
derive the $\mathcal{G}_\M$ {\em global momentum} of a given power system.
To give an example-driven insight into the approach we are proposing, we
present a simple case-study whose first goal is to show
how \RefEq{E:HalfSysNoise2SF} can be derived.
\subsection{Low-frequency characterization}
The large-signal model of the system in \RefFig{F:twomac}, in which the $b_1(t)$ and $b_2(t)$ small-signals are neglected,
is given by the following equations:
\begin{equation}
\begin{array}{rcl}
\dot{\delta}_1 &=& \Omega \left( \omega_1 - \omega_0 \right) \\
\dot{\delta}_2 &=& \Omega \left( \omega_2 - \omega_0 \right) \\
\M_1 \dot{\omega}_1 &=& P_{m_1} - P_{e_1}(\delta_1,\delta_2) - \D_1 \left( \omega_1 - \omega_0 \right) \\
\M_2 \dot{\omega}_2 &=& P_{m_2} - P_{e_2}(\delta_1,\delta_2) - \D_2 \left( \omega_2 - \omega_0 \right) \\
T_{g_1} \dot P_{m_1} &=& P_{{m_{\mathrm{eq}}}_1}-P_{m_1}-{k_{g_1}}{R_1^{-1}}\left( \omega_1 - \omega_0 \right)\\
\end{array}~,
\label{E:Eq1}
\end{equation}
where
\begin{equation*}
\begin{array}{lcl}
P_{e_1}(\delta_1,\delta_2) &\!\!\!\!\!=\!\!\!\!\!&\frac{1}{R^2+X^2} \left[
    R \rho_1^2 - 
    R \rho_1 \rho_2 \cos(\delta_1 - \delta_2) \right. + \\
    &\!\!\!\!\! \!\!\!\!\!& +\left. X \rho_1 \rho_2 \sin(\delta_1 - \delta_2)
\right] + \frac{1}{R_L} \rho_1^2~, \\
\end{array}
\end{equation*}
and
\begin{equation*}
\begin{array}{lcl}
P_{e_2}(\delta_1,\delta_2) &\!\!\!\!\!=\!\!\!\!\!& \frac{1}{R^2+X^2} \left[
    R \rho_2^2 - 
    R \rho_1 \rho_2 \cos(\delta_2 - \delta_1) + \right. \\[3mm]
    &\!\!\!\!\! \!\!\!\!\!& \left. X \rho_1 \rho_2 \sin(\delta_2 - \delta_1) \right]~.
\end{array}
\end{equation*}
%We assumed that the turbine governor is model implements a single (dominant)
%pole transfer function.
The last equation in \RefE{E:Eq1} models the turbine governor connected to 
the $G_1$ generator\footnote{We modeled turbine governors with a
dominant pole transfer function, as done for example in \cite{Sajadi2022}.
Some turbine governors can require a more complex transfer function
consisting of a zero and a pair of higher frequency poles. To keep notation simple, we adopted in this example the former modeling. Nonetheless, the proposed methodology is compatible with any governor model.}.
The power system is assumed to be at steady state, with 
${\omega_k=\omega_{\mathrm{eq}_k}=\omega_0=1}$,
${\delta_k=\delta_{\mathrm{eq}_k}}$, and 
${P_{m_k} = P_{{m_{\mathrm{eq}}}_k} = P_{e_k}(\delta_{\mathrm{eq}_1},
\delta_{\mathrm{eq}_2})}$ for $k\in\{1,2\}$. 
The small signal equivalent model of \RefE{E:Eq1}, including now the small-signal additive 
perturbations projected according to \RefE{E:DecompNoise}, is
\begin{equation}
\begin{array}{rcl}
\Delta \dot \delta_1 &\!\!\!\!\!=\!\!\!\!\!&\Omega \Delta {\omega_1} \\
\Delta \dot \delta_2 &\!\!\!\!\! =\!\!\!\!\! &\Omega \Delta {\omega_2} \\
\M_1 \Delta \dot \omega_1 &\!\!\!\!\! =\!\!\!\!\! &\xi
        (\Delta \delta_2 - \Delta \delta_1) -
\D_1 \Delta \omega_1 +\Delta P_{m_1}+ \M_1 b_{{\delta\omega}_1}(t)\\
\M_2 \Delta \dot \omega_2 &\!\!\!\!\! =\!\!\!\!\! &\xi
        (\Delta \delta_1 - \Delta \delta_2) -
\D_2 \Delta {\omega_2} + \M_2 b_{{\delta\omega}_2}(t) \\
T_{g_1} \Delta \dot P_{m_1} &\!\!\!\!\! =\!\!\!\!\! & -\Delta P_{m_1} -k_{g_1}{R_1}^{-1}\Delta \omega_1\\
\end{array}
\label{E:Eq4}
\end{equation}
where 
\begin{equation*}
\xi = \frac{\rho_1 \rho_2}{R^2+X^2} \left(R 
\sin(\delta_{\mathrm{eq}_1} - \delta_{\mathrm{eq}_2}) + X  \cos(\delta_{\mathrm{eq}_1} - \delta_{\mathrm{eq}_2}) \right)~.
\end{equation*}

By transforming \RefEq{E:Eq4} in the Laplace domain,
the $\Delta {\omega_1}(s)$ and $\Delta {\omega_2}(s)$ small-signal
variations of the rotor speeds of the two synchronous generators can be
written as
\begin{equation}
\begin{array}{l}
\displaystyle{\Delta {\omega_1}(s) = 
\frac{\beta_3^1(s)s^3+\beta_2^1(s)s^2+\beta_1^1(s)s+\beta_0^1(s)}
     {\alpha_4 s^4 + \alpha_3 s^3 + \alpha_2 s^2 + \alpha_1 s + \alpha_0}} \\[5mm]
\displaystyle{\Delta {\omega_2}(s) = 
\frac{\beta_3^2(s)s^3+\beta_2^2(s)s^2+\beta_1^2(s)s+\beta_0^2(s)}
     {\alpha_4 s^4 + \alpha_3 s^3 + \alpha_2 s^2 + \alpha_1 s + \alpha_0}} 
\end{array}~,
\label{E:Eq6}
\end{equation}
where the $\alpha _k$ and $\beta_k^j$ coefficients are reported in Appendix B.
For {\em sufficiently small} values of $s$ we neglect in \RefE{E:Eq6} the 
terms $\alpha _k s^k$ for $k\in\{3,4\}$, and $\beta_k^j s^k$ for 
$k\in\{2,3\}$ and $j\in\{1, 2\}$. 
Furthermore, in the coefficients reported in Appendix B we neglect all the 
terms divided by $\xi \Omega$ since we assume $\xi \Omega \gg 1$. 
Thus we approximate \RefE{E:Eq6} as reported 
in \RefE{E:Eq7} (next page). 
The same expression can be straightforwardly derived from the most generic one in \RefE{E:HalfSysNoise2SF}.

\begin{figure}[t]
\begin{center}
\includegraphics[scale=0.86]{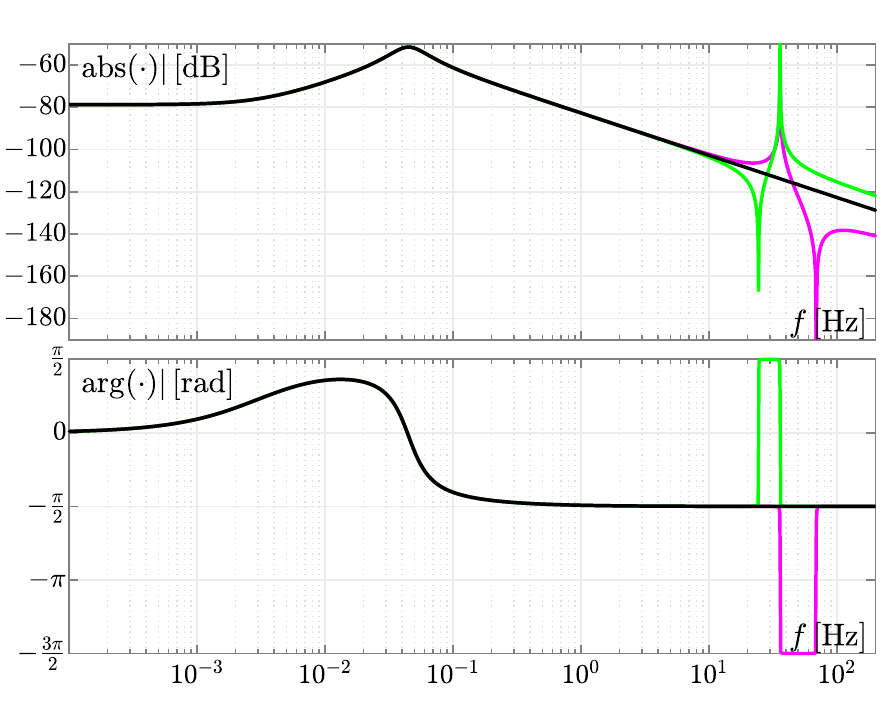}
\end{center}
\caption{The modulus and the phase of $\Delta \omega_1$ (in magenta, see \RefE{E:Eq6}), 
$\Delta \omega_2$ (in green, see \RefE{E:Eq6}), and $\Delta \omega$ (in black, see \RefE{E:Eq7}) are shown in 
the upper and lower panel, respectively. 
These curves are derived for the power system in \RefFig{F:twomac} by assuming
$S_{B_1} = S_{B_2} = 100\,\mega\volt\ampere$, $D_1=D_2=1$, $H_1=4\,\second$, 
$H_2=2.5\,\second$, $T_{g_1}=50\,\second$, ${k_{g_1}}{R_1^{-1}}=50\,S_{B_1}$, 
$V_{\mathrm{base}}=100\,\kilo\volt$, $Z_{\mathrm{base}}=100\,\ohm$, 
$R=X=0.001\,Z_{\mathrm{base}}$, and $R_L=\nicefrac{Z_{\mathrm{base}}}{0.9}$. The power flow solution provides $P_{{m_{\mathrm{eq}}}_1}=40\,\mega\watt$ and 
$P_{{m_{\mathrm{eq}}}_2}=50\,\mega\watt$.
$\M_1b_{{\delta\omega}_1}(t)$ and $\M_2b_{{\delta\omega}_2}(t)$ are 
sinusoidal functions whose phase is zero and whose amplitude is fixed at $100\,\kilo\watt$ and 
$500\,\kilo\watt$, respectively, for each value of $f$.
\label{F:twomac_resp}}
\end{figure}

\InRefFig{F:twomac_resp} shows the plots corresponding to 
the expressions in \RefE{E:Eq6} and \RefE{E:Eq7}.
As it can be noticed, the approximation is extremely good for $f<10\hertz$. The expressions of $\Delta {\omega_1}(s)$ and
$\Delta {\omega_2}(s)$ at low frequencies turn out to be {\em identical}; this means that 
the $\Delta {\delta_1}$ and $\Delta {\delta_2}$ corresponding 
angles vary in a synchronized way.
%%%%%%%%%%% BEGIN FIGURE %%%%%%%%%%%%
%%%%%%%%%%% END  FIGURE %%%%%%%%%%%%

%%%%%%%%%%%%%%%%%%%%%%%%%%%%%%%%%%%%%%%%%%%%%%%%%%%%%%%%%%%%%%%%%%%%%%%%%%%%%%%
\begin{figure*}[t]
\normalsize
\begin{equation}
\displaystyle{\Delta {\omega}_1(s)\simeq\Delta {\omega}_2(s)\simeq\Delta {\omega}(s) \equiv 
\frac{(sT_{g_1}+1)(\M_1 b^{\mathrm{lf}}_{{\delta\omega}_1}(s)+\M_2 b^{\mathrm{lf}}_{{\delta\omega}_2}(s))}{s^2 \left(\M_1+\M_2\right) T_{g_1}+s \left((\D_1+\D_2)T_{g_1}+\M_1+\M_2\right)+\left(\left(\D_1+\D_2\right) +{k_{g_1}R_1^{-1}}\right)}}
\label{E:Eq7}
\end{equation}
\hrulefill
\vspace*{4pt}
\end{figure*}
%%%%%%%%%%%%%%%%%%%%%%%%%%%%%%%%%%%%%%%%%%%%%%%%%%%%%%%%%%%%%%%%%%%%%%%%%%%%%%%

\subsection{Global momentum estimation}
\label{S:experiment}
\InRefEq{E:Eq7} can be recast as
\begin{equation}
\begin{array}{l c l}
\Delta \omega_1(s)& = & \left(\frac{c_1}{s-a_1} + \frac{c_2}{s-a_2}\right)\eta(s)\\[5mm]
& = &\frac{(c_1+c_2)s - c_1 a_2 - c_2 a_1}{s^2 - s(a_1+a_2) + a_1 a_2}\eta(s)\\[5mm]
& = &\frac{\left(\frac{1}{\M_1+\M_2} s + \frac{1}{(\M_1+\M_2) T_{g1}}\right)\eta(s)}
  {s^2 + \left( \frac{1}{T_{g1}} + \frac{\D_1+\D_2}{\M_1+\M_2}\right)s + 
  \frac{\D_1 + \D_2 + {k_{g_1}R_1^{-1}}}{(\M_1+\M_2) T_{g1}}} 
\end{array}
\label{E:TwoTg1}
\end{equation}
where $a_1$ and $a_2$ are the poles, $c_1$ and $c_2$ are the residues, and
$\eta(s) = \M_1 b^{\mathrm{lf}}_{{\delta\omega}_1}(s)+\M_2 b^{\mathrm{lf}}_{{\delta\omega}_2}(s)$.

From \RefE{E:TwoTg1} we see that $\mathcal{G}_M=\M_1+\M_2$
can be derived through the $c_1 + c_2$ term by performing an {\em experiment}.
Let us assume to vary $\M_1$ of a known constant value $\Delta \M$ 
(varying $\M_2$ is totally equivalent). This variation does not alter the power flow solution of the system but how the $G_1$ generator reacts to variations of its rotor speed.
Before the experiment, we have $c_1 + c_2 = (\M_1+\M_2)^{-1}$
and after the experiment
$\widehat{c}_1 + \widehat{c}_2 = (\M_1 + \M_2 + \Delta \M)^{-1}$
from which we derive
\begin{equation}
\mathcal G_M = \frac{\widehat{c}_1 + \widehat{c}_2}
  {c_1 + c_2 - \widehat{c}_1 - \widehat{c}_2 }\Delta \M ~.
\label{E:GlobM1}
\end{equation}
At first we perform a proper frequency scan of the system by acting on 
either $b_1 (t)$ or $b_2(t)$ (or both) to numerically derive \RefE{E:Eq7}. 
Then, we use the \ac{vf} technique \cite{772353,1645204,Triverio} to fit \RefE{E:TwoTg1} and determine the pairs of $c_1$, $c_2$ and $\widehat{c}_1$ and $\widehat{c}_2$ 
coefficients (i.e., before and after the experiment) and, thus, $\mathcal G_M$.

This approach presents several drawbacks. First and foremost, in
practice it is typically not possible to act on the inertia constant of
the synchronous machines. Secondly, a complete frequency scan is also
unfeasible. Lastly, the rotor speed of the synchronous machines
is unlikely to be available. In the following we show how to overcome
these issues.

\section{A method for the estimation of the power-system global momentum}
\label{S:method}
The {\em experiment} described in \RefSec{S:experiment} can be implemented
by resorting to a \acf{gf} \acf{cig}. 
This device may already be present in the power system or
alternatively can be inserted with the specific purpose of allowing
the estimation of the power system global momentum.
As shown in \RefSec{S:NumExp}, the state equations governing the 
dynamics of this device are similar to those reported in \RefE{E:FullSys}. 
Hence its equivalent rotor speed ${\omega}_{\mathrm{gf}}$ becomes one of the entries of the 
$\vec \omega$ vector and its equivalent momentum constant one of the 
diagonal elements of the $\vec \M$ matrix. 
Proper low-frequency small-signal tones injected through the \ac{gf} 
\ac{cig} itself may contribute to the 
$\vec \eye_{1,N}\vec \M\vec b_{\delta\omega}(s)$ term in 
\RefEq{E:HalfSysNoise2SF}. 
By observing the frequency response of the equivalent rotor speed of 
the \ac{gf} \ac{cig}, once those tones are injected, we can obtain a proper 
set of samples of \RefE{E:HalfSysNoise2SF}. 
We recall that \RefE{E:HalfSysNoise2SF} provides the expression in the 
Laplace domain shared (at low frequency) by all the components of
$\vec \omega$ and hence of ${\omega}_{\mathrm{gf}}$. By varying the \ac{gf} \ac{cig} equivalent momentum constant and fitting 
the samples of  \RefE{E:HalfSysNoise2SF} before and after this variation, 
it is possible to derive the {\em global momentum} of the entire power system.

We underline that the reader must not be confused at this point: by 
injecting small-signal power perturbations in the \ac{cig} and measuring its 
virtual rotor speed, {\em we do not estimate the \ac{cig} virtual 
contribution to the overall system  momentum}, as it is done in several 
papers in the literature. 
On the contrary, we estimate the {\em global momentum} of the entire 
power system (\ac{cig} included). In other words, the \ac{cig} can be viewed as
a probing-signal source, whose virtual inertia is known and can be modified during the experiment needed to
obtain $\mathcal G_M$.

The method we developed gives a continuous estimate of the global momentum.
Since we use the \ac{vf} algorithm \cite{772353, 1645204, Triverio} 
to estimate the residues and the poles of \RefE{E:HalfSysNoise2SF}, 
we need frequency samples of both its modulus and phase.
This forbids resorting to the power spectral density of ${\omega}_{\mathrm{gf}}$
when its fluctuation is solely given by the noisy generated/absorbed powers.
We thus opted to modulate the power injected by the \ac{cig} by 
a discrete set of $\mathcal N_{\mathcal T}$ deterministic and coherent 
small-signal sinusoidal tones $s_w(t)$ ($w=1,\dots,\mathcal N_{\mathcal T}$).
The choices of $\mathcal N_{\mathcal T}$ and of the $f_w$ frequency of each tone
depend on the bandwidth that has to be explored to fit 
\RefE{E:HalfSysNoise2SF}. 
It is worth mentioning that $\mathcal N_{\mathcal T}$ must be fixed in 
excess with respect to the residues and poles number of 
\RefE{E:HalfSysNoise2SF}. 
Injection of small-signal tones in a power system is not an unusual technique
and it was already adopted in \cite{9099871,7913675}.
A good description of this technique is in \cite{5338001}.
The $s_w(t)$ tones are very slowly varying and of modest magnitude, therefore
they do not impact the stability of the power system.

The equivalent inertia constant of the \ac{cig} varies as a square waveform 
of amplitude $\Delta \M$, period $T_{\Delta \M}$, and duty cycle $50\%$. 
The power system is stimulated by injecting the $s_w(t)$ waveforms and 
the time samples of ${\omega}_{\mathrm{gf}}$ are collected.
%Since in a real power system all
%electrical and mechanical quantities are affected by noise, we filtered ${\omega}_{\mathrm{gf}}$ by resorting to a proper bandpass filter. 
We computed the $\gamma_{w_c}$ and $\gamma_{w_s}$ direct and quadrature 
components of the Fourier integrals of 
${\omega}_{\mathrm{gf}}$ at each $f_q$ frequency as
\begin{equation}
\begin{array}{l}
\displaystyle{\gamma_{w_c} = \mu f_w 
  \int_{t_0}^{t_0+\frac{\mu}{f_w}}{{\omega}_{\mathrm{gf}}(t)
  \cos(2\pi \mu f_w t ) dt}} \\[4mm]
\displaystyle{\gamma_{w_s} = \mu f_w 
  \int_{t_0}^{t_0+\frac{\mu}{f_w}}{{\omega}_{\mathrm{gf}}(t)
  \sin(2\pi \mu f_w t ) dt}}   ~.
\end{array}
\label{E:FI}
\end{equation}
The  $\gamma_{w_c}$ and $\gamma_{w_s}$ terms constitute the frequency samples that feed the \ac{vf} algorithm. 
Note that if it is necessary to reduce the effects of noise to increase
the signal-to-noise ratio, the integrals 
in \RefE{E:FI} can be computed over a time interval $\nicefrac{\mu}{f_w}$ 
which is a multiple of the period of the corresponding tone $s_w(t)$. 
The $t_0$ time instant in \RefE{E:FI} coincides with the rising and 
falling edges of the square waveform used to periodically change the virtual 
inertia of the \ac{cig}.
Assuming $s_1(t)$ as the tone with the lowest frequency, $f_1$ suggests 
how to choose $T_{\Delta \M}$, i.e.,
$\nicefrac{T_{\Delta \M}}{2} > \nicefrac{\mu}{f_1}$.

As shown in the next section, the proposed estimation method can be applied 
to bigger power systems than that shown in \RefFig{F:twomac} by resorting to 
\RefE{E:HalfSysNoise2SF}. Indeed, also \RefE{E:HalfSysNoise2SF} can be recast 
through a partial fraction decomposition analogous to \RefE{E:TwoTg1}. 
Thus, even in more complex cases, the process to estimate the global 
momentum of inertia still relies on \RefE{E:GlobM1} and \acl{vf}.

\section{Numerical examples}
\label{S:NumExp}
\subsection{Virtual synchronous generator}
A \acf{gf} \acf{cig} implements a control scheme that simulates 
the mechanical dynamical behavior of a synchronous machine (swing equation)
by means of a power converter.
The goal is to provide inertia, damping, primary frequency control, and 
voltage control to a network with a significant penetration of renewable 
energy sources and therefore reduced inertia. 

The proposed approach to estimate the $\mathcal G_M$ global momentum exploits the \ac{cig} model described in \cite{9625933}.
It provides virtual inertia by implementing the swing equation with 
frequency droop control and replicates the stator impedance of the 
synchronous generator.
Importantly, the vast majority of \acp{cig} that provide synthetic
inertia reproduce exclusively the mechanical behavior
of a synchronous machine (i.e., the swing equation), while failing to replicate
its full electro-mechanical behavior due to windings, including dampers. 
On the other hand, they add the dynamics of the electronic converter.
The \ac{cig} model in \cite{9625933} belongs to this category and therefore 
implements only the 
swing equation, leading to a different spectral footprint at high frequencies
with respect to a real synchronous generator. The swing equation implemented in \cite{9625933} is
\begin{equation*}
\begin{array}{rcl}
\dot{\delta}_{\mathrm{gf}} &=& \Omega \left( \omega_{\mathrm{gf}} - \omega_{0} 
   \right) \\[1mm]
T_a \dot{\omega}_{\mathrm{gf}} &=& \left(1+\eta_{\mathrm{gf}}\right) P_g - P_e - 
   K_d \left( \omega_{\mathrm{pll}} - \omega_{\mathrm{gf}} \right) + \\
   & & - K_w \left( \omega_{\mathrm{gf}} - 
   \omega_{\mathrm{0}} \right) \\[1mm]
P_e &=& \frac{V_d I_d + V_q I_q}{P_{\mathrm{ref}}}
\end{array}~,
\end{equation*}
where
$V_d$, $I_d$, $V_q$, and $I_q$ are the voltages and currents of the generator in
the dq-frame that lead to the $P_e$ electrical active power (per-unit), 
$P_{\mathrm{ref}}$ is the power rating of the \ac{cig}, 
$T_a$ is the (virtual) inertia constant,
$P_g$ is the generated power setpoint at system frequency $\omega_0$
(per-unit), 
$K_w$ is the load-damping, 
$\delta_{\mathrm{gf}}$ if the angle deviation of the virtual rotor,
$\omega_{\mathrm{gf}}$ is the virtual rotor angular speed (per-unit),
$\omega_{\mathrm{pll}}$ is the electrical angular frequency of the voltage
at the connection bus (per-unit),
$K_d$ regulates the virtual rotor speed according to the
electrical angular frequency of the bus to which the \ac{cig} is connected, thus
emulating frequency slip (in our case $K_d=0$). 
$\eta_{\mathrm{gf}}$ is the small signal used to perturb the
power system, which modulates the active power setpoint of the \ac{cig}.

\subsection{Load models}
\label{S:NOISELOADS}
The loads that are present in the power systems considered here as case studies allow to perturb their active power. The $l-$th of these loads (for $l=1,\dots,L$) is modeled as
\begin{equation}
  \mathrm{L}_l = \left( 1 + \eta_l(t) \right) \, P_{{\mathrm{L0}}_l} \left(
  \frac{\left|V_l\right|}{V_{0_l}} \right)^{\gamma}~,
\label{E:LoadModel}
\end{equation}
where  $P_{{\mathrm{L0}}_l}$ is the nominal active
power of the load, $V_{0_l}$ is the load voltage rating, $V_l$ is
the bus voltage at which the load is connected, and $\gamma$ governs
the dependence of the load on bus voltage (hereafter assumed to be null).
By applying the $\eta_l(t)$ small signal we can perturb the load power.

In the time domain simulations of stochastic load fluctuations, we
assume that $\eta_l(t)$ is an \acf{ou} process \cite{Milano2013}.
The \ac{ou} processes (one for each load) are defined through the set 
of \acp{sde}
\begin{equation}
  \label{E:eta}
  d \vec \eta = -\vec \Upsilon \vec \eta \, dt + \vec \Sigma \, d \vec W_t ~,
\end{equation}
where the drift ${\vec \Upsilon \in \R^{L \times L}}$ and diffusion
${\vec \Sigma\in \R^{L \times L}}$ are diagonal matrices with positive
entries, $\vec W_t \in \R^L$ is a vector of Wiener processes, and
the differentials rather than time derivatives are utilized to account
for the idiosyncrasies of \acp{sde}.
The \ac{ou} processes are characterized by a mean-reversion property and
exhibit bounded standard deviation.  Moreover, these processes show a
spectrum that is an accurate model of the stochastic variability
of power loads \cite{HIRPARA2015409, NWANKPA1990338, Milano2013,
7540898, 8447491}.

To carry out the simulations discussed below, the numerical
integration of the multi-dimensional \ac{ou} process in
\RefE{E:eta} was based on the numerical scheme proposed by
Gillespie in \cite{Gillespie1996}.  Furthermore, the second-order
trapezoidal implicit weak scheme for stochastic differential equations
with colored noise, available in the simulator \textsc{pan}
\cite{ngcas, Bizzarri201451,Linaro2022}, was adopted \cite{Milshtein1994}.

\subsection{The \textsc{ieee 39-bus} test system}
\begin{figure}[tb]
\begin{center}
  \includegraphics[width=0.95\columnwidth]{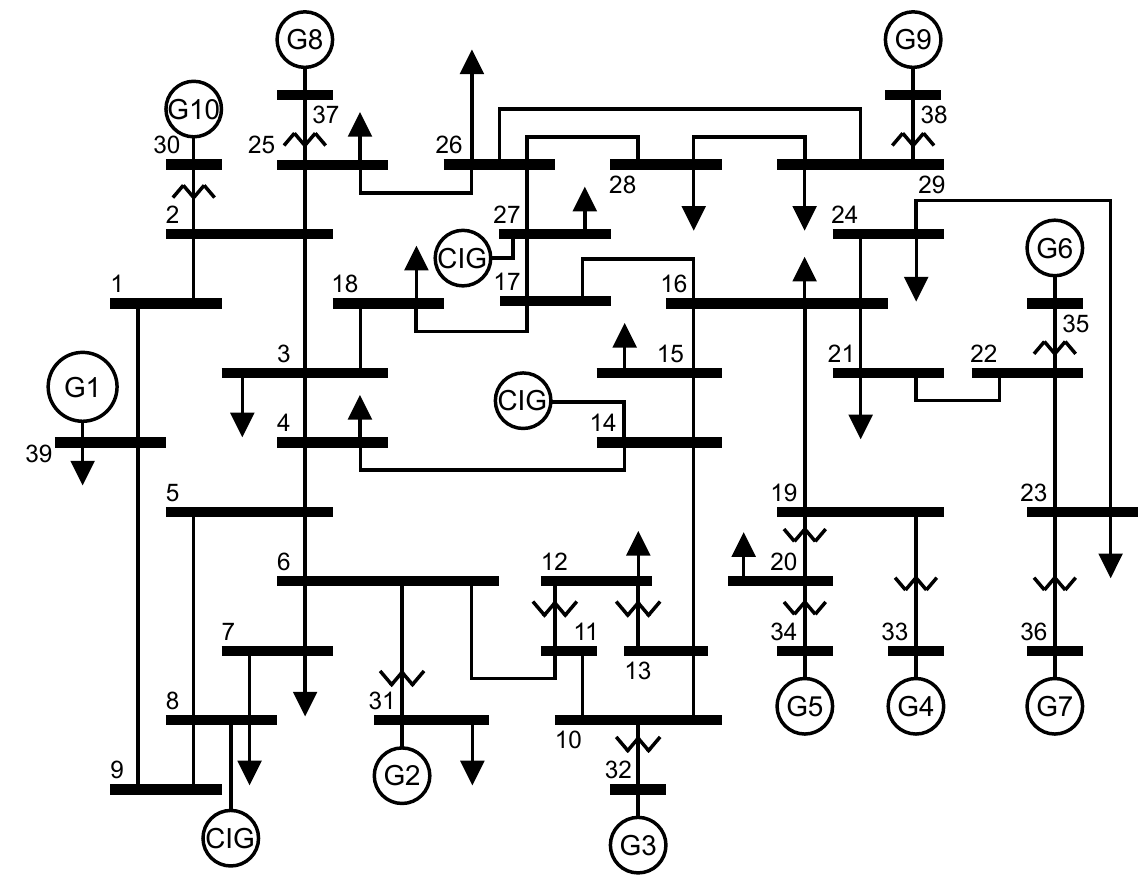}
\end{center}
\caption{Schematic of the \textsc{ieee 39-bus} system. 
\label{F:ieee39Sch}}
\end{figure}

We used as first benchmark the \textsc{ieee 39-bus} system \cite{4113518}.
The grid contains $10$ generators and $46$ lines and it is a simplified model 
of the New England power system.
Its schematic is reported in \RefFig{F:ieee39Sch}.
The $G_1$ generator models the aggregate behavior of a large number 
of generators.
This is reflected in its momentum value 
($\mathcal M_{G_1}=100\,\giga\joule$), which is one order of magnitude larger 
than that of the other generators in the network (see Table \ref{T:PRATH}).
\begin{table}[b]
  \caption{Synchronous generators $H$ and $S_{B}$}
  \centering
      \begin{tabular}{ c | c | c | c | c | c}
      \toprule
      Gen. & $H\,[\second]$& $S_{B}\,[\mega\watt]$ &
      Gen.	& $H\,[\second]$& $S_{B}\,[\mega\watt]$\\
      \midrule
      $G_{1}$ & $5.00$  & $10000$ & $G_{6}$ & $4.35$ & $800$ \\
      $G_{2}$ & $4.33$  & $700$ & $G_{7}$ & $3.77$ & $700$\\
      $G_{3}$ & $4.47$  & $800$ & $G_{8}$ & $3.47$ & $700$\\
      $G_{4}$& $3.57$& $800$& $G_{9}$& $3.45$& $1000$\\
      $G_{5}$& $4.33$& $600$& $G_{10}$& $4.20$& $1000$\\
      \bottomrule
    \end{tabular}
    \label{T:PRATH}
\end{table}
The \textsc{ieee 39-bus} system version we started from is that in the 
distribution of \textsc{powerfactory} by \textsc{digsilent}.

The \ac{cig}, used to generate the perturbing power tones and to perform 
the {\em experiment} in the proposed global momentum estimation, is connected 
at \textsc{bus14}.
There is not a preferred bus to which the \ac{cig} should be connected.
We inserted two additional \acp{cig} at \textsc{bus8} and \textsc{bus27}
to show that the proposed estimation algorithm takes into account every
component affecting the global momentum. 
The former models an aggregated wind power plant and the latter 
a battery storage system.
Each \ac{cig} is characterized by a $5\,\giga\joule$ momentum.

\begin{figure}[b]
\begin{center}
  \includegraphics[scale=0.9]{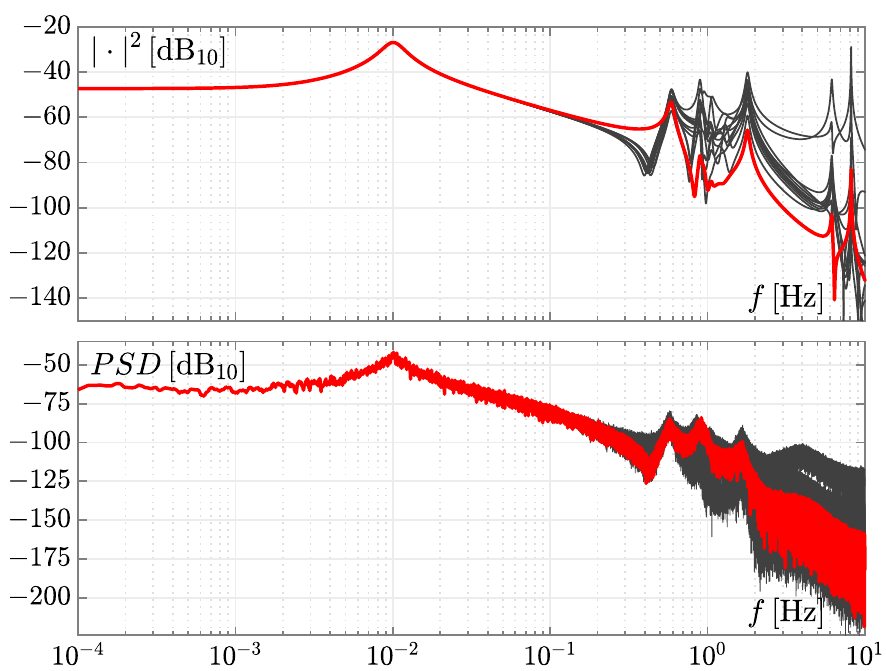}
\end{center}
\caption{Upper panel: the magnitude squared of the transfer functions between $\eta_{\mathrm{gf}}(t)$ and the rotor speed of each synchronous machine and \ac{cig} of the (modified) \textsc{ieee 39-bus} system. 
Lower panel: power spectral densities of the rotor speed of each 
synchronous machine and \ac{cig} of the (modified) \textsc{ieee 39-bus} 
system when the stochastic noise in the loads is turned on. 
In both the panels the red curve refers to the \ac{cig} connected 
at \textsc{bus14}.
\label{F:Ieee39AcScan}}
\end{figure}

Firstly, we performed an (ideal) frequency scan assuming $\eta_{\mathrm{gf}}(t)$ as a small-signal sinusoidal source and by turning off the stochastic variations of the loads. So doing, we computed all the transfer functions between $\eta_{\mathrm{gf}}(t)$ and the rotor speed of each synchronous machine and \ac{cig}.
Figure \ref{F:Ieee39AcScan} (upper panel) reports these 
transfer functions.
We can notice that they almost perfectly overlap at frequencies below $0.1\,\hertz$
while they sensibly differ for higher frequency values.
This result carries similar information as that in \RefFig{F:twomac_resp}.
Note that these transfer functions show how rotor speeds deviate from 
$\omega_0$ under the assumption of small signal behavior (linear behavior
in the neighborhood of the equilibrium point, i.e., power-flow).
These spectra show that if we think of a step perturbation, during the 
first time
interval after its application, say 
${100\,\milli\second \rightarrow 10\,\hertz}$, the synchronous generators 
lose synchrony and counteract power imbalance in a non-coordinated way.
This behavior persists up to $2\,\second \rightarrow 0.5\,\hertz$, at
which point the synchronous generators and \acp{cig} go toward 
re-synchronization (low frequency behavior).
By observing the low frequency overlapping of all the curves
in \RefFig{F:Ieee39AcScan}, the global momentum can be determined 
through \RefE{E:HalfSysNoise2SF} by fitting {\em only} the curve related to 
the \ac{cig} connected at \textsc{bus14} in the $[6,30]\,\milli\hertz$ 
frequency interval, since all rotor speeds are described by the same behavior 
in this frequency interval (principal frequency system dynamics), 
as predicted by our analysis.
The frequency behavior in this band is exclusively due to the 
mechanical characteristics of the power system contributed by synchronous 
generators and prime movers (swing equation) and by \acp{cig} that 
implement virtual inertia contribution.

\begin{figure}[tb]
\begin{center}
  \includegraphics[scale=1.0]{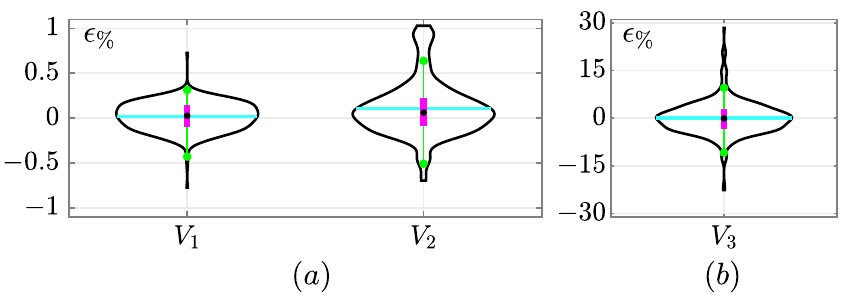} \\
\end{center}
\caption{
The horizontal cyan segments and the black solid circle markers
correspond to the mean $\mu$ and the median $\mu_{\text{dn}}$ of the
results, respectively. The magenta bars represent the \ac{iqr},
viz. the spread difference between the $75^{\mathrm{th}}$ and
$25^{\mathrm{th}}$ percentiles of the data. The green solid circle
markers represent the upper adjacent value (i.e., the largest
observation that is less than or equal to the third quartile plus
$1.5 \times \ac{iqr}$) and the lower adjacent value (i.e., the
smallest observation that is greater than or equal to the first
quartile minus $1.5 \times \ac{iqr}$).
Panel (a). The $V_1$ violin plot ($\mu=0.019$, $\mu_{\text{dn}}=0.026$, and $\ac{iqr}=0.232$) refers to the results obtained by
exploiting an ideal frequency scan assuming $\eta_{\mathrm{gf}}(t)$ as a small-signal sinusoidal source.
The simulations were carried out in the frequency-domain by resorting to the AC analysis.
The $V_2$ violin plot ($\mu=0.105$, $\mu_{\text{dn}}=0.062$, and $\ac{iqr}=0.302$) refers to the results obtained by injecting 
a set of $\mathcal N_{\mathcal T} = 10$ deterministic and coherent small signal sinusoidal tones
through the \ac{cig} connected at \textsc{bus14}. The simulations were carried out in the time-domain and
the integrals in \RefEq{E:FI} were computed to feed the \ac{vf} algorithm.
Both $V_1$ and $V_2$ were obtained by assuming that all the loads as noiseless.
Panel (b). Contrary to $V_2$, the $V_3$ violin plot ($\mu=0.074$, $\mu_{\text{dn}}=-0.104$, and $\ac{iqr}=5.48$) was derived by turning on the stochastic noise in the loads.
\label{F:Ieee39ScanMomentum}}
\end{figure}

We exploited this frequency scan to estimate the global momentum of 
the \textsc{ieee 39-bus} system.
We performed 500 independent frequency scans and for each of those the inertia 
constant of each generator was uniformly randomly varied by $\pm 30\%$ 
with respect to its nominal value.
This was done to test the method on a large set of inertia 
configurations\footnote{Some of these configurations can lead to
(almost) identical global momentum values but with different partitions 
among each synchronous generator and \ac{cig}.}.
At each run, we used the \ac{vf} method to fit the frequency behavior of the \ac{cig} virtual rotor angular frequency before and after the 
experiment and estimated the global momentum.
The left violin plot in \RefFig{F:Ieee39ScanMomentum}(a) summarizes
the performance of the proposed approach, in terms of the
$\epsilon_\%$ percent relative error, in estimating the global
momentum for each one of the $500$ random configurations\footnote{A
  violin plot is a combination of a box plot and a kernel  density
  plot: specifically, it starts with a box plot and then adds a
  rotated kernel density plot to each side of the box plot.}.

Since as said before, this ideal frequency scan is not practical, we
turned on the injection of the discrete set of $\mathcal N_{\mathcal T} = 10$ 
deterministic and coherent small signal sinusoidal tones by the \ac{cig}
connected at \textsc{bus14}.
We performed $500$ time domain simulations lasting 
$15\,\minute$ and we randomly varied the inertia configuration.
Before varying the inertia configuration as detailed in Section \ref{S:method}
we performed the ``experiment'' and computed the global momentum.
The information of the $\epsilon_\%$ percent relative error in estimating the global momentum 
are shown by the right violin plot in \RefFig{F:Ieee39ScanMomentum}(a).
We can notice that the overall performance is worsened even if it remains very good since the relative accuracy in determining the global momentum is in any case lower than $2.0\%$.

Finally we turned on the stochastic noise in the loads. 
In the lower panel of \RefFig{F:Ieee39AcScan}, we report the frequency 
behavior of the rotor speed of all the synchronous machines and 
\acp{cig} when all the loads of the network are perturbed as described 
in \RefSec{S:NOISELOADS}. 
We used $L=19$ independent $\eta_z$ small-signal stochastic noise sources, 
one for each power load.
We choose a $2\,\second$ reversion time and set $\sigma_z$ (see \RefE{E:eta}) in such a way that the standard deviation of $\eta_l(t)$ is $0.5\%$ of $P_{L0_{l}}$ (nominal load active power) in \RefE{E:LoadModel}. 
The zero mean implies that the stochastic loads power fluctuations do not perturb, on average, the operating point of the system.
By observing the spectral densities in \RefFig{F:Ieee39AcScan} (lower panel),
we notice a different behavior of the rotor speed deviations at 
frequency above $0.1\,\hertz$ but once more spectra almost perfectly 
overlap at lower frequencies. 
This means that the effect of the stochastic noise sources will be 
superimposed to that of the $\mathcal N_\T$ tones injected by the \ac{cig}.

We ran $500$ time-domain large-signal simulations with the nominal momentum 
configuration
and with a magnitude of the injected tones that cause a peak power variation
less than $2.5\%$ of the nominal power of the \textsc{ieee 39-bus} system.
The violin plot giving information about the relative error in the
global momentum estimation is shown in \RefFig{F:Ieee39ScanMomentum}(b).

We see that the relative error is further increased with respect to the
previous cases.
This is due to the decreased \ac{snr} that makes more difficult the 
fitting of equation \RefE{E:HalfSysNoise2SF} by the \ac{vf} method.

\subsection{The \textsc{ieee118} power system}
As a second benchmark we used the \textsc{ieee 118-bus} system.
It represents an approximation of the American Electric Power system 
(in the U.S. Midwest) as of December 1962.
It contains $19$ generators, $35$ synchronous condensers, $177$ lines, 
$9$ transformers, and $91$ loads.
The original model is available in the distribution of \textsc{matpower} 
\cite{Zimmerman:2010}, but it does not contain any dynamic model.
There are several versions enhanced with dynamic 
models; guiding rules can be found in \cite{7553421}.

The \ac{gf} \ac{cig}, used to generate the perturbing power tones and
to perform the experiment, is connected at \textsc{bus38}.
As already said for the \textsc{ieee 39-bus} there is not a preferred bus
to which the \ac{cig} should be connected. 

\begin{figure}[!!!t]
\begin{center}
  \includegraphics[scale=0.9]{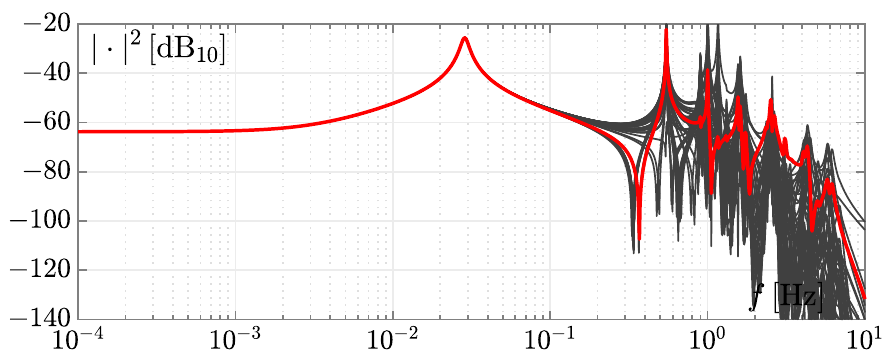}
\end{center}
\caption{The magnitude squared of the transfer functions between $\eta_{\mathrm{gf}}(t)$ injected through the \ac{gf} \ac{cig} connected at \textsc{bus101} and the rotor speed of each synchronous machine and \ac{cig} of the (modified) \textsc{ieee 118-bus} system. The red curve refers to the (virtual) rotor speed of the \ac{cig} connected at \textsc{bus101}.
\label{F:Ieee118AcScan}}
\end{figure}

The result by the (ideal) frequency scan assuming
$\eta_\mathrm{gf}(t)$ as a small-signal sinusoidal source is shown in 
\RefFig{F:Ieee118AcScan}.
We see that as for the \textsc{ieee 39-bus} system all the angular
frequencies of synchronous machine rotors and \ac{cig} overlap almost
perfectly at low frequency {($< 100\,\milli\hertz$)}.

\begin{figure}[!!!t]
\begin{center}
  \includegraphics[scale=1.0]{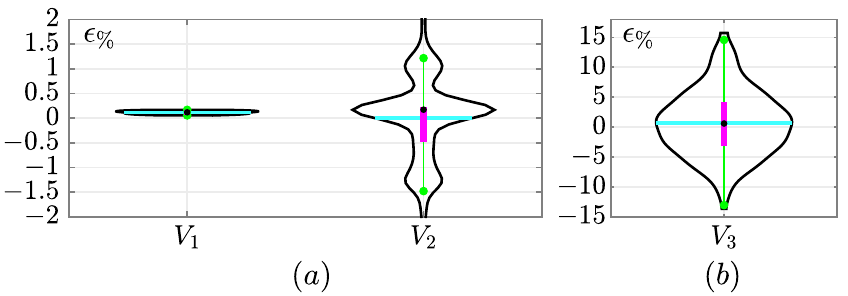} \\
\end{center}
\caption{
{Panel (a). The $V_1$ violin plot: $\mu=0.115$, $\mu_{\text{dn}}=0.117$, and $\ac{iqr}=0.055$.
The $V_2$ violin plot: $\mu=-0.002$, $\mu_{\text{dn}}=0.170$, and $\ac{iqr}=0.680$.
Panel (b). The $V_3$ violin plot: $\mu= 0.675$, $\mu_{\text{dn}}=0.592$, and $\ac{iqr}=7.01$. The simulation setup for the $V_k$ violin plot ($k\in\{1,2,3\}$) is the same one of the $k-$th plot in \RefFig{F:Ieee39ScanMomentum}.}
\label{F:Ieee118ScanMomentum}}
\end{figure}

We thus repeated the same simulation we had done with the \textsc{ieee39}, 
that is, the inertia of each generator was uniformly randomly varied
of $\pm 30\%$ with respect to its nominal value and we estimated the global
inertia by a frequency scan in the time domain with noiseless loads 
and finally with noisy loads.
The results are summarized by the violin plots in 
\RefFig{F:Ieee118ScanMomentum} and have the same meaning of those in 
\RefFig{F:Ieee39ScanMomentum}.
We see that very good estimation results of global momentum are obtained 
also for the \textsc{ieee 118-bus} system.

\section{Concluding remarks}
We propose a method to estimate the global momentum of electrical power
systems that requires only measured data from a \acf{gf} \acf{cig} that 
is already present in the system or that can be added for this purpose.

The approach is fully data-driven and proved accurate even in complex 
systems and can take into account both conventional and virtual 
synchronous machines. 

Estimations of global momentum of the \textsc{ieee 39-bus} and larger 
\textsc{ieee 118-bus} test systems lead to an average percent relative error lower than 
$1\%$ (with an interquartile range lower than $7\%$),  when the inertia of each generator is
varied in a uniformly random fashion by $\pm 30\%$ and noise by loads is considered.
The estimation algorithm runs in less than $10\,\second$ on a conventional
computer.

We believe that the proposed technique can be useful to monitor the total 
frequency support provided by conventional and virtual-synchronous devices. 
In the near future we plan to explore other extrapolation procedures that allow
the use of noise by stochastic load/generation and to use this technique 
on real data measured by \textsc{pmu}.

\appendices
\section{}
To derive the $\vec v_1$ left-eigenvector reported in \RefE{E:v1} we write
\begin{equation}
\vec A^{\T}=\left[\!\!\!\!
  \begin{array}{c c c}
    \vec \oo_N & -\vec \Pi \vec \M^{-1} & \vec \oo_N\\
    \Omega \vec \eye_N & \vec \M^{-1} \vec \D & -\vec T^{-1}_{g} \vec k_{g} \vec R^{-1} \\
	\vec \oo_N & \vec \M^{-1}& -\vec T^{-1}_{g}\\
	\end{array}
	\!\!\!\!\right]
\label{E:AT}
\end{equation}
and we have to solve $\vec A^{\T} \vec v_1=\vec \oo_{3N,1}$. By choosing
$\vec v_1 = (\vec v_\delta^{\T},\vec v_\omega^{\T},\vec v_{P_m}^{\T})^{\T}$, this implies
\begin{equation*}
\left\{
\begin{array}{l}
\vec \Pi \vec \M^{-1} \vec v_{\omega} = \vec \oo_{N,1}\\
\Omega \vec v_{\delta}+\vec \M^{-1} \vec \D \vec v_{\omega} -\vec T^{-1}_{g} \vec k_{g} \vec R^{-1} \vec v_{P_m} = \vec \oo_{N,1}\\
\vec \M^{-1}\vec v_{\omega} -\vec T^{-1}_{g} \vec v_{P_m} = \vec \oo_{N,1}\\
\end{array}\
\right. ~.
\label{E:v1eqs}
\end{equation*}
From the first equation above, being $\ker(\vec \Pi)=\vec \eye_{N,1}$,
we find $\vec v_{\omega}=\vec \M \vec \eye_{N,1}$. Hence, from the third equation,
$\vec v_{P_m} = \vec T_{g} \eye_{N,1}$, and, from the second one,
$\vec v_{\delta}=\Omega^{-1}(\vec \D+\vec k_{g} \vec R_{g}^{-1})\vec \eye_{N,1}$.
Since $\vec u_1=[\vec \eye_{1,N}, \vec \oo_{1,2N}]^{\T}$, then
\begin{equation*}
\vec v_1^{\T} \vec u_1 = \Omega^{-1}\eye_{1,N}\left(\vec \D+\vec k_{g} \vec R_{g}^{-1}\right)\vec \eye_{N,1}~,
\end{equation*}
and it is possible to normalize $\vec v_1^{\T}$ in such a way that $\vec v_1^{\T} \vec u_1=1$ thus obtaining \RefEq{E:v1}.

\section{}
\begin{align*}
\label{E:coeff}
\beta_3^1(s) =& \M_1 \M_2 R_1 T_{g_1} b_{{\delta\omega}_1}(s)\\
\beta_2^1(s) =& \M_1 \left(\D_2 T_{g_1}+\M_2\right) R_1 b_{{\delta\omega}_1}(s)\\
\beta_1^1(s) =& \left(\rule{0mm}{4.7mm}(b_{{\delta\omega}_1}(s) \M_1+b_{{\delta\omega}_2}(s) \M_2)T_{g_1} +\frac{}{}\right.\\
& + \left.\frac{b_{{\delta\omega}_1}(s)\M_1\D_2}{\xi \Omega}\right)\xi \Omega R_1\\
\beta_0^1(s) =& \left(b_{{\delta\omega}_1}(s) \M_1+b_{{\delta\omega}_2}(s) \M_2\right)\xi \Omega R_1\\
\beta_3^2(s) =& \M_1 \M_2 R_1 T_{g_1} b_{{\delta\omega}_2}(s)\\
\beta_2^2(s) =& \M_2 \left(\D_1 T_{g_1}+\M_1\right)R_1 b_{{\delta\omega}_2}(s) \\
\beta_1^2(s) =& \left(\rule{0mm}{4.4mm}(b_{{\delta\omega}_1}(s) \M_1+b_{{\delta\omega}_2}(s) \M_2) T_{g_1} + \frac{}{}\right.\\
& +\left.\frac{b_{{\delta\omega}_2}(s)\M_2(\D_1+k_{g_1}R_1^{-1})}{\xi\Omega}\right)\xi \Omega R_1\\
\beta_0^2(s) =& \beta_0^1(s)\\
\alpha_4 =&  \M_1 \M_2 R_1 T_{g_1}\\
\alpha_3 =& R_1(T_{g_1}(D_ 2 M_ 1 + D_ 1 M_ 2) + 
   M_ 1 M_ 2)\\
\alpha_2 =& \left(\frac{\D_2 \left(\D_1T_{g_1}+\M_1\right)+\D_1\M_2}{\xi \Omega} + \frac{}{}\right.\\
 & \left.+\left(\M_1+\M_2\right) T_{g_1} +\frac{k_1 R_1^{-1} \M_2}{\xi \Omega}\right)\xi  \Omega R_1\\
\alpha_1 =& \left(\rule{0mm}{6mm}(\D_1 + \D_2) T_{g_1} + \M_1+ \M_2 + \frac{}{}\right.\\
&+ \left.\frac{\D_2 \left(\D_1+k_{g_1}R_1^{-1}\right)}{\xi \Omega}\!\!\right)\xi \Omega R_1\\
\alpha_0 =& (D_ 1 + D_ 2  + k_{g_1} R_1^{-1})\xi\Omega R_ 1\\
\end{align*}
% ----- ----- ----- ----- ----- ----- ----- ----- -----
%\bibliographystyle{IEEEtran}
%\bibliography{biblio}

\begin{thebibliography}{10}
	\providecommand{\url}[1]{#1}
	\csname url@rmstyle\endcsname
	\providecommand{\newblock}{\relax}
	\providecommand{\bibinfo}[2]{#2}
	\providecommand\BIBentrySTDinterwordspacing{\spaceskip=0pt\relax}
	\providecommand\BIBentryALTinterwordstretchfactor{4}
	\providecommand\BIBentryALTinterwordspacing{\spaceskip=\fontdimen2\font plus
		\BIBentryALTinterwordstretchfactor\fontdimen3\font minus
		\fontdimen4\font\relax}
	\providecommand\BIBforeignlanguage[2]{{%
			\expandafter\ifx\csname l@#1\endcsname\relax
			\typeout{** WARNING: IEEEtran.bst: No hyphenation pattern has been}%
			\typeout{** loaded for the language `#1'. Using the pattern for}%
			\typeout{** the default language instead.}%
			\else
			\language=\csname l@#1\endcsname
			\fi
			#2}}
	
	\bibitem{KONTIS:2021}
	E.~O. Kontis, I.~D. Pasiopoulou, D.~A. Kirykos, T.~A. Papadopoulos, and G.~K.
	Papagiannis, ``Estimation of power system inertia: A comparative assessment
	of measurement-based techniques,'' \emph{Electr. Pow. Syst. Res.}, vol. 196,
	p. 107250, 2021.
	
	\bibitem{PRABHAKAR:2022}
	K.~Prabhakar, S.~K. Jain, and P.~K. Padhy, ``Inertia estimation in modern power
	system: A comprehensive review,'' \emph{Electr. Pow. Syst. Res.}, vol. 211,
	p. 108222, 2022.
	
	\bibitem{5338001}
	J.~W. Pierre, N.~Zhou, F.~K. Tuffner, J.~F. Hauer, D.~J. Trudnowski, and W.~A.
	Mittelstadt, ``Probing signal design for power system identification,''
	\emph{{IEEE} Trans. Power Syst.}, vol.~25, no.~2, pp. 835--843, 2010.
	
	\bibitem{Zhou:2006}
	N.~Zhou, J.~Pierre, and J.~Hauer, ``Initial results in power system
	identification from injected probing signals using a subspace method,''
	\emph{{IEEE} Trans. Power Syst.}, vol.~21, no.~3, pp. 1296--1302, 2006.
	
	\bibitem{Chakraborty:2022}
	R.~Chakraborty, H.~Jain, and G.-S. Seo, ``A review of active probing-based
	system identification techniques with applications in power systems,''
	\emph{International Journal of Electric Power and Energy Systems}, vol. 140,
	p. 108008, 2022.
	
	\bibitem{Hosaka:2019}
	N.~Hosaka, B.~Berry, and S.~Miyazaki, ``The world's first small power
	modulation injection approach for inertia estimation and demonstration in the
	island grid,'' in \emph{2019 8th International Conference on Renewable Energy
		Research and Applications (ICRERA)}, 2019, pp. 722--726.
	
	\bibitem{Tamrakar:2020}
	U.~Tamrakar, N.~Guruwacharya, N.~Bhujel, F.~Wilches-Bernal, T.~M. Hansen, and
	R.~Tonkoski, ``Inertia estimation in power systems using energy storage and
	system identification techniques,'' in \emph{2020 International Symposium on
		Power Electronics, Electrical Drives, Automation and Motion (SPEEDAM)}, 2020,
	pp. 577--582.
	
	\bibitem{Rauniyar:2021}
	M.~Rauniyar, S.~Berg, S.~Subedi, T.~M. Hansen, R.~Fourney, R.~Tonkoski, and
	U.~Tamrakar, ``Evaluation of probing signals for implementing moving horizon
	inertia estimation in microgrids,'' in \emph{2020 52nd North American Power
		Symposium, NAPS 2020}, 2021.
	
	\bibitem{7913675}
	J.~Zhang and H.~Xu, ``{Online Identification of Power System Equivalent Inertia
		Constant},'' \emph{{IEEE} Trans. Ind. Electron.}, vol.~64, no.~10, pp.
	8098--8107, Oct. 2017.
	
	\bibitem{Sajadi2022}
	A.~Sajadi, R.~W. Kenyon, and B.-M. Hodge, ``Synchronization in electric power
	networks with inherent heterogeneity up to 100\% inverter-based renewable
	generation,'' \emph{Nature Communications}, vol.~13, no.~1, 2022.
	
	\bibitem{Gorjao2022}
	L.~Rydin~Gorjão, L.~Vanfretti, D.~Witthaut, C.~Beck, and B.~Schäfer, ``Phase
	and amplitude synchronization in power-grid frequency fluctuations in the
	nordic grid,'' \emph{IEEE Access}, vol.~10, pp. 18\,065--18\,073, 2022.
	
	\bibitem{Guo2021}
	Y.~Guo, D.~Zhang, Z.~Li, Q.~Wang, and D.~Yu, ``Overviews on the applications of
	the kuramoto model in modern power system analysis,'' \emph{International
		Journal of Electric Power and Energy Systems}, vol. 129, 2021.
	
	\bibitem{Zhu2018}
	L.~Zhu and D.~J. Hill, ``Stability analysis of power systems: A network
	synchronization perspective,'' \emph{SIAM Journal on Control and
		Optimization}, vol.~56, no.~3, p. 1640 – 1664, 2018.
	
	\bibitem{Nishikawa2015}
	T.~Nishikawa, F.~Molnar, and A.~E. Motter, ``Stability landscape of power-grid
	synchronization,'' \emph{IFAC-PapersOnLine}, vol.~48, no.~18, pp. 1--6, 2015,
	4th IFAC Conference on Analysis and Control of Chaotic Systems CHAOS 2015.
	
	\bibitem{Motter2013}
	A.~E. Motter, S.~A. Myers, M.~Anghel, and T.~Nishikawa, ``Spontaneous synchrony
	in power-grid networks,'' \emph{Nature Physics}, vol.~9, no.~3, p. 191 –
	197, 2013.
	
	\bibitem{Ramaswamy1995}
	G.~Ramaswamy, G.~Verghese, L.~Rouco, C.~Vialas, and C.~DeMarco, ``Synchrony,
	aggregation, and multi-area eigenanalysis,'' \emph{IEEE Transactions on Power
		Systems}, vol.~10, no.~4, pp. 1986--1993, 1995.
	
	\bibitem{Kundur1994}
	P.~Kundur, N.~Balu, and M.~Lauby, \emph{Power system stability and control},
	ser. EPRI power system engineering series.\hskip 1em plus 0.5em minus
	0.4em\relax McGraw-Hill, 1994.
	
	\bibitem{9625933}
	B.~Barać, M.~Krpan, T.~Capuder, and I.~Kuzle, ``Modeling and initialization of
	a virtual synchronous machine for power system fundamental frequency
	simulations,'' \emph{IEEE Access}, vol.~9, pp. 160\,116--160\,134, 2021.
	
	\bibitem{772353}
	B.~Gustavsen and A.~Semlyen, ``Rational approximation of frequency domain
	responses by vector fitting,'' \emph{{IEEE} Trans. Power Del.}, vol.~14,
	no.~3, pp. 1052--1061, 1999.
	
	\bibitem{1645204}
	B.~Gustavsen, ``Improving the pole relocating properties of vector fitting,''
	\emph{{IEEE} Trans. Power Del.}, vol.~21, no.~3, pp. 1587--1592, 2006.
	
	\bibitem{Triverio}
	P.~Triverio, \emph{Vector Fitting, in Handbook on Model Order Reduction}.\hskip
	1em plus 0.5em minus 0.4em\relax De Gruyter, Berlin, 2021.
	
	\bibitem{Farkas1994}
	M.~Farkas, \emph{Periodic motions}.\hskip 1em plus 0.5em minus 0.4em\relax
	Springer-Verlag, 1994.
	
	\bibitem{Kuznetsov2004}
	Y.~A. Kuznetsov, \emph{Elements of Applied Bifurcation Theory}, 3rd~ed.\hskip
	1em plus 0.5em minus 0.4em\relax Springer-Verlag, 2004.
	
	\bibitem{Demir2000}
	A.~Demir, A.~Mehrotra, and J.~Roychowdhury, ``Phase noise in oscillators: a
	unifying theory and numerical methods for characterization,'' \emph{{IEEE}
		Trans. Circuits Syst. {I}}, vol.~47, no.~5, pp. 655--674, 2000.
	
	\bibitem{Aulbach1984}
	B.~Aulbach, \emph{Continuous and Discrete Dynamics Near Manifolds of
		Equilibria}, ser. Lecture Notes in Mathematics.\hskip 1em plus 0.5em minus
	0.4em\relax Springer-Verlag, 1984.
	
	\bibitem{Sauer1990}
	P.~Sauer and M.~A. Pai, ``Power system steady-state stability and the load-flow
	jacobian,'' \emph{{IEEE} Trans. Power Syst.}, vol.~5, no.~4, pp. 1374--1383,
	Nov. 1990.
	
	\bibitem{Machowski2020}
	J.~Machowski, Z.~Lubosny, J.~Bialek, and J.~Bumby, \emph{Power System Dynamics:
		Stability and Control}.\hskip 1em plus 0.5em minus 0.4em\relax Wiley, 2020.
	
	\bibitem{Nwankpa1990}
	C.~Nwankpa and S.~Shahidehpour, ``Colored noise modelling in the reliability
	evaluation of electric power systems,'' \emph{Appl. Math. Model.}, vol.~14,
	no.~7, pp. 338--351, 1990.
	
	\bibitem{Milano2013}
	F.~{Milano} and R.~{Zárate-Miñano}, ``A systematic method to model power
	systems as stochastic differential algebraic equations,'' \emph{{IEEE} Trans.
		Power Syst.}, vol.~28, no.~4, pp. 4537--4544, Nov. 2013.
	
	\bibitem{Hirpara2015}
	R.~H. Hirpara and S.~N. Sharma, ``An {O}rnstein-{U}hlenbeck process-driven
	power system dynamics,'' \emph{IFAC-PapersOnLine}, vol.~48, no.~30, pp.
	409--414, 2015, 9th IFAC Symposium on Control of Power and Energy Systems
	CPES 2015.
	
	\bibitem{7540898}
	C.~Roberts, E.~M. Stewart, and F.~Milano, ``Validation of the
	ornstein-uhlenbeck process for load modeling based on {$\mu$PMU}
	measurements,'' in \emph{2016 Power Systems Computation Conference (PSCC)},
	2016, pp. 1--7.
	
	\bibitem{Abundo2013}
	M.~Abundo, ``On the representation of an integrated gauss-markov process,''
	\emph{Scientiae Mathematicae Japonicae Online E-2013}, p. 719 – 723, 2013.
	
	\bibitem{9099871}
	F.~Zeng, J.~Zhang, G.~Chen, Z.~Wu, S.~Huang, and Y.~Liang, ``Online estimation
	of power system inertia constant under normal operating conditions,''
	\emph{IEEE Access}, vol.~8, pp. 101\,426--101\,436, 2020.
	
	\bibitem{HIRPARA2015409}
	R.~H. Hirpara and S.~N. Sharma, ``An {O}rnstein-{U}hlenbeck process-driven
	power system dynamics,'' \emph{IFAC-PapersOnLine}, vol.~48, no.~30, pp.
	409--414, 2015, 9th IFAC Symposium on Control of Power and Energy Systems
	CPES 2015.
	
	\bibitem{NWANKPA1990338}
	C.~Nwankpa and S.~Shahidehpour, ``Colored noise modelling in the reliability
	evaluation of electric power systems,'' \emph{Appl. Math. Model.}, vol.~14,
	no.~7, pp. 338--351, 1990.
	
	\bibitem{8447491}
	H.~Hua, Y.~Qin, C.~Hao, and J.~Cao, ``{Stochastic Optimal Control for Energy
		Internet: A Bottom-Up Energy Management Approach},'' \emph{{IEEE} Trans. Ind.
		Informat.}, vol.~15, no.~3, pp. 1788--1797, Mar. 2019.
	
	\bibitem{Gillespie1996}
	D.~T. Gillespie, ``Exact numerical simulation of the ornstein-uhlenbeck process
	and its integral,'' \emph{Phys. Rev. E}, vol.~54, pp. 2084--2091, Aug 1996.
	
	\bibitem{ngcas}
	F.~Bizzarri and A.~Brambilla, ``{PAN} and {MPanSuite}: Simulation vehicles
	towards the analysis and design of heterogeneous mixed electrical systems,''
	in \emph{NGCAS, Genova, Italy}, Sept. 2017, pp. 1--4.
	
	\bibitem{Bizzarri201451}
	F.~Bizzarri, A.~Brambilla, G.~{Storti Gajani}, and S.~Banerjee, ``Simulation of
	real world circuits: Extending conventional analysis methods to circuits
	described by heterogeneous languages,'' \emph{{IEEE} Circuits Syst. Mag.},
	vol.~14, no.~4, pp. 51--70, 2014.
	
	\bibitem{Linaro2022}
	D.~Linaro, D.~del Giudice, F.~Bizzarri, and A.~Brambilla, ``Pansuite: A free
	simulation environment for the analysis of hybrid electrical power systems,''
	\emph{Electr. Pow. Syst. Res.}, vol. 212, 2022.
	
	\bibitem{Milshtein1994}
	G.~N. Milshtein and M.~V. Tret'yakov, ``Numerical solution of differential
	equations with colored noise,'' \emph{Journal of Statistical Physics},
	vol.~77, no.~3, pp. 691--715, 1994.
	
	\bibitem{4113518}
	T.~Athay, R.~Podmore, and S.~Virmani, ``{A Practical Method for the Direct
		Analysis of Transient Stability},'' \emph{{IEEE} Trans. Power App. Syst.},
	vol. PAS-98, no.~2, pp. 573--584, Mar./Apr. 1979.
	
	\bibitem{Zimmerman:2010}
	R.~D. Zimmerman, C.~E. Murillo-S{\'a}nchez, and R.~J. Thomas, ``{MATPOWER}:
	{S}teady-state operations, planning, and analysis tools for power systems
	research and education,'' \emph{{IEEE} Trans. Power Syst.}, vol.~26, no.~1,
	pp. 12--19, Feb. 2010.
	
	\bibitem{7553421}
	``Ieee recommended practice for excitation system models for power system
	stability studies,'' \emph{IEEE Std 421.5-2016 (Revision of IEEE Std
		421.5-2005)}, pp. 1--207, 2016.
	
\end{thebibliography}
% ----- ----- ----- ----- ----- ----- ----- ----- -----

\end{document}